\documentclass{jfm}
\usepackage{graphicx}
\usepackage{newtxtext}
\usepackage{newtxmath}
\usepackage{natbib}
\usepackage{hyperref}
\hypersetup{
    colorlinks = true,
    urlcolor   = blue,
    citecolor  = blue,
}

\newcommand{\RomanNumeralCaps}[1]
\nolinenumbers

\makeatletter

\usepackage{multirow}

\title[3D Buoyant HF: constant release]{Three-dimensional buoyant hydraulic fracture growth: constant release from a point source}
\author[A. M\"ori, B. Lecampion]{Andreas M\"ori and Brice Lecampion\thanks{Email for correspondence: brice.lecampion@epfl.ch, andreas.mori@epfl.ch} } 
\affiliation{Geo-Energy Laboratory - Gaznat Chair on Geo-Energy,\\  
Ecole Polytechnique F\'ed\'erale de Lausanne, \\
 ENAC-IIC-GEL-EPFL,  Station 18, CH-1015, Switzerland}

\pubyear{2022} 
\volume{--} 
\pagerange{--} 
\date{received May 02, 2022; revised September 01, 2022, accepted -}  

\makeatother

\usepackage{babel}

\begin{document}
\title{Three-dimensional buoyant hydraulic fractures: constant release from
a point source}
\maketitle
\begin{abstract}
Hydraulic fractures propagating at depth are subjected to buoyant
forces caused by the density contrast between fluid and solid. This
paper is concerned with the analysis of the transition from an initially
radial towards an elongated buoyant growth -- a critical topic for
understanding the extent of vertical hydraulic fractures in the upper
Earth crust. Using fully coupled numerical simulations and scaling
arguments, we show that a single dimensionless number governs buoyant
hydraulic fracture growth: the dimensionless viscosity of a radial
hydraulic fracture at the time when buoyancy becomes of order one.
It quantifies if the transition to buoyancy occurs when the growth
of the radial hydraulic fracture is either still in the regime dominated
by viscous flow dissipation or is already in the regime where fracture
energy dissipation dominates. A family of fracture shapes emerge at
late time from finger-like (toughness regime) to inverted elongated
cudgel-like (viscous regime). 3D toughness dominated buoyant fractures
exhibit a finger-like shape with a constant volume toughness dominated
head and a viscous tail having a constant uniform horizontal breadth:
there is no further horizontal growth past the onset of buoyancy.
However, if the transition to buoyancy occurs while in the viscosity
dominated regime, both vertical and horizontal growths continue to match
scaling arguments. As soon as the fracture toughness is not strictly
zero, horizontal growth stops when the dimensionless horizontal toughness
becomes of order one. The horizontal breadth follows the predicted
scaling.
% -- and compares fairly well with existing toughness-dominated
%experiments.
\end{abstract}

\section{Introduction}

We investigate the propagation of three-dimensional hydraulic fractures
emerging from a point source accounting for buoyancy forces. Hydraulic
fractures (HF) are tensile fluid-filled fractures propagating under
internal fluid pressure which exceed the minimum compressive in-situ
stress of the surrounding media \citep{Deto16}. HFs are encountered
in various engineering applications \citep{SmMo15,JeCh13,GeMu10}
but also occur in nature due to fluid over-pressure at depth, for example
during the formation of magmatic intrusions \citep{RiTa15,SpSh87,LiKe91}.
The minimum physical ingredients to model HF growth are lubrication
flow within the elastically deformable fracture coupled to quasi-static
fracture propagation under the assumption of linear elastic fracture
mechanics (LEFM) \citep{Deto16}. In the absence of buoyancy, theoretical
predictions reproduce well experiments in brittle and impermeable
materials \citep{BuDe08,LeDe17,XiYo17}.

HF propagate radially from a point source and remain so in the absence
of buoyancy. For such a geometry, the growth is initially dominated
by energy dissipation in viscous flow and transitions to a regime
dominated by fracture energy dissipation at late time (in association
with the increase of the fracture perimeter). Growth solutions in
both regimes are well known \citep{AbKe76,SpSh85,SaDe02}. The presence
of buoyant forces necessarily elongates the fracture. A large body
of work investigated the impact of buoyant forces on two-dimensional
plane strain fractures \citep{Weer71,SpTu90,SpSh87,List90,RoLi07,SpTu85}.
The early work of \citet{Weer71} focused on a toughness-dominated
fracture with a linear pressure gradient and did not consider any
fluid flow. These considerations lead to a fluid-filled pocket with
a stress intensity factor equal to the material resistance at the
upper tip, respectively zero at the lower tip of such a bubble crack.
A two-dimensional pulse is hence created. Owing to the lack of coupling
with lubrication flow, a description of the dynamics of its ascent
is missing. A first attempt to include viscous effects was done by
\citet{SpSh87} and \citet{SpTu90}. \citet{List90} has obtained
solutions as a function of a dimensionless fracture toughness with
a focus on small fracture toughness / large viscosity cases. These
2D buoyant HFs exhibit a distinct head region, close to the propagating
edge, where a hydrostatic gradient develops and a tail region where
viscous flow occurs within a conduit of constant width. The solution
in the so-called toughness dominated regime was obtained by \citet{RoLi07}
complementing earlier work \citep{List90,LiKe91}.

A pseudo-three-dimensional solution for viscosity-dominated buoyant
fractures was developed by \citet{List90b} in conjunction with a
scaling analysis. Assuming a large aspect ratio for the fracture allows
for a partial uncoupling of elasticity and lubrication flow. The boundary
conditions of his model are such that the fracture has an unprescribed
open upper end, such that this approximate solution is deemed to
be valid in the near-source region. It predicts an ever-increasing
horizontal extent of the fracture, which must be limited in the case
of a finite, non-zero fracture toughness. A planar three-dimensional
solution has been derived by \citet{GaGe22} \citep[see also][]{GaGe14,GeGa14} in the limit of large
material toughness. This approximate solution is constructed by matching
a constant breadth (blade-like) viscosity-dominated tail with a 3D
toughness-dominated head under a hydrostatic gradient. This approximate
toughness solution shows a propagating head akin to a constant 3D
Weertmann pulse \citep{Weer71} propagating upward due to the linear
extension of a fixed breadth in a viscosity-dominated tail. Recently
the problem of a finite volume release has been investigated in the
limit of zero fluid viscosity numerically by \citet{DaRi20}, focusing
on the minimal volume required for the start of buoyant propagation.
Similar simulations are reported in \citet{SaZi20}, where lubrication
flow is included but only small volume releases are investigated without
an extensive study of the late-time growth of buoyant 3D HF.

In this contribution, we investigate the transition of initially radial expansion
HFs to the late-time fully three-dimensional buoyant regimes accounting
for the complete coupling between elastohydrodynamic lubrication flow
and linear elastic fracture mechanics. We notably aim to clarify the
domain of validity of previous contributions in the viscosity and
toughness dominated limits and fully understand the solution space
of three-dimensional buoyant fractures under constant volume release.

\section{Formulation and methods\label{sec:ProbFormNMethod}}

\begin{figure}
\begin{center}
\includegraphics[width=0.95\textwidth]{Fig1_Sketch.pdf}
\end{center}
\caption{Schematic of a buoyancy-driven hydraulic fracture (head $\to$ red,
tail $\to$ green, source region $\to$ grey). The tail length is
reduced for illustration, indicated by dashed lines and a shaded area.
The fracture propagates in the $x\vert z$ plane with a gravity vector
${\displaystyle \mathbf{g}}$ oriented in ${\displaystyle -z}$. The
fracture front ${\displaystyle \mathcal{C}\left(t\right)}$, fracture
surface ${\displaystyle \mathcal{A}\left(t\right)}$ (dark gray area),
opening ${\displaystyle w\left(x,z,t\right)}$, net pressure ${\displaystyle p\left(x,z,t\right)}$,
the local normal velocity of the fracture ${\displaystyle v_{c}\left(x_{c},z_{c}\right)}$
with ${\displaystyle \left({\displaystyle x_{c},}z_{c}\right)\in{\displaystyle \mathcal{C}\left(t\right)}}$
characterize fracture growth under a constant release rate ${\displaystyle Q_{o}}$
in a medium with a linear confining stress with depth ${\displaystyle \sigma_{o}\left(z\right)}$.
${\displaystyle \ell^{\textrm{head}}\left(t\right)}$ and ${\displaystyle b^{\textrm{head}}\left(t\right)}$
denote the length and breadth of the head, ${\displaystyle \ell\left(t\right)}$
is the total fracture length, and ${\displaystyle b\left(z,t\right)}$
is the local breadth of the fracture.}
\label{fig:Sketch}
\end{figure}

\subsection{Mathematical formulation\label{subsec:S2:MathForm}}

We consider a pure opening mode (mode I) hydraulic fracture propagating
from a point source located at depth in the $x\vert z$ plane as sketched
in figure \ref{fig:Sketch}. This $x\vert z$ plane is perpendicular
to the minimum in-situ stress $\sigma_{o}(z)$ (taken positive in
compression). We assume that the minimum in-situ stress acts in the
$y$-direction and is thus perpendicular to the gravity vector ${\displaystyle \mathbf{g}=\left(0,0,-g\right)}$
(with ${\displaystyle g}$ the earth's gravitational acceleration).
Owing to the possibly large fracture dimensions, we account for a
linear vertical gradient of the in-situ stress (resulting from the
initial solid equilibrium). Assuming a linear elastic medium with
uniform properties, the quasi-static balance of momentum for a planar
tensile hydraulic fracture reduces to a hyper-singular boundary integral
equation over the fracture surface $\mathcal{A}(t)$. This integral
equation relates the fracture width ${\displaystyle w\left(x,z,t\right)}$
to the net loading, which is equivalent to the difference between
the fluid pressure inside the fracture ${\displaystyle p_{f}\left(x,z,t\right)}$
and the minimum compressive in-situ stress ${\displaystyle \sigma_{o}\left(x,z\right)}$
\citep{CrSt83,HiKe96}
\begin{equation}
p\left(x,z,t\right)=p_{f}\left(x,z,t\right)-\sigma_{o}\left(x,z\right)=-\frac{E^{\prime}}{8\pi}\int_{\mathcal{A}(t)}\frac{w\left(x^{\prime},z^{\prime},t\right)}{\left[\left(x^{\prime}-x\right)^{2}+\left(z^{\prime}-z\right)^{2}\right]^{3/2}}\text{\text{d}\ensuremath{x^{\prime}}d}z^{\prime}\label{eq:S2:Elasticity}
\end{equation}
where $E^{\prime}=E/\left(1-\nu^{2}\right)$ is the plane-strain modulus
with $E$ the material Young's modulus and $\nu$ its Poisson's ratio.
As typically observed in the Earth's crust \citep{HeMo18,Corn15,JaCo07},
the minimum confining stress ${\displaystyle \sigma_{o}\left(z\right)}$
increases linearly with depth proportional to the solid weight ${\displaystyle \gamma_{s}=\rho_{s}g}$
multiplied by a dimensionless lateral earth pressure coefficient ${\displaystyle \alpha}$.
Accounting for the downward orientation of the gravitational vector
in the chosen coordinate system (see figure \ref{fig:Sketch}), the
vertical gradient for ${\displaystyle \sigma_{o}\left(z\right)}$
is linear over the entire medium
\begin{equation}
\text{d}\sigma_{o}\left(z\right)/\text{d}z=-\alpha\rho_{s}g\to\mathbf{\nabla}\sigma_{o}=\alpha\rho_{s}\mathbf{g}.
\label{eq:S2:GradSigo}
\end{equation}
Fluid flow within the thin deforming fracture is governed by lubrication
theory \citep{Batc67}. Neglecting any fluid exchange between the
rock and the fracture (a reasonable assumption for tight formations
and high viscosity fluids), the width-averaged continuity equation
for an incompressible fluid reduces to
\begin{equation}
\frac{\partial w\left(x,z,t\right)}{\partial t}+\nabla\cdot\left(w\left(x,z,t\right)\mathbf{v}_{f}\left(x,z,t\right)\right)=\delta(x)\delta(z)Q_{o}(t)\label{eq:S2:IncompFluidContinuity}
\end{equation}
where $\mathbf{v}_{f}\left(x,z\right)$ is the width averaged fluid
velocity, and $Q_{o}$ is the volumetric flow rate at the point source
located at the origin ${\displaystyle \left(x,z\right)=\left(0,0\right)}$.
Additionally, the assumption of no fluid exchange with the surrounding
medium dictates that the total volume of the fracture is equal to
the total volume released. Assuming a constant release rate $Q_{o}$,
the global volume conservation is chiefly:
\begin{equation}
\mathcal{V}\left(t\right)=\int_{\mathcal{A}(t)}w\left(x,z\right)\text{d}x\text{d}z=Q_{o}t.\label{eq:2SBala}
\end{equation}
Assuming laminar flow and a Newtonian rheology, the fluid flux ${\displaystyle \mathbf{q}\left(x,z,t\right)=w\left(x,z,t\right)\mathbf{v}_{f}\left(x,z,t\right)}$
reduces to Poiseuille's law accounting for buoyancy forces:
\begin{equation}
\mathbf{q}\left(x,z,t\right)=w\left(x,z,t\right)\mathbf{v}_{f}\left(x,z,t\right)=-\frac{w\left(x,z,t\right)^{3}}{\mu^{\prime}}\left(\mathbf{\nabla}p_{f}\left(x,z,t\right)-\rho_{f}\mathbf{g}\right)\label{eq:S2:Lubr}
\end{equation}
where $\mu^{\prime}=12\mu_{f}$ is the equivalent parallel plates
fluid viscosity, $\mu_{f}$ is the fluid viscosity, and $\rho_{f}$
is the fluid density. Introducing the net pressure ${\displaystyle p\left(x,z,t\right)=p_{f}\left(x,z,t\right)-\sigma_{o}\left(z\right)}$
and using equation (\ref{eq:S2:GradSigo}), (\ref{eq:S2:Lubr}) is
rewritten as
\begin{equation}
\mathbf{q}\left(x,z,t\right)=-\frac{w\left(x,z,t\right)^{3}}{\mu^{\prime}}\left(\mathbf{\nabla}p\left(x,z,t\right)+\varDelta\gamma\frac{\mathbf{g}}{\left|\mathbf{g}\right|}\right)\label{eq:S2:FluidFlow}
\end{equation}
where ${\displaystyle \varDelta\gamma=\varDelta\rho g=\left(\alpha\rho_{s}-\rho_{f}\right)g}$
is the effective buoyancy contrast of the system. For a value of $\alpha = 1$, it equals the buoyancy contrast between the solid and the fluid. Values of the lateral earth pressure coefficient ${\displaystyle \alpha}$ different than one, have
no other influence than affecting the value of the effective buoyancy
contrast ${\displaystyle \varDelta\gamma}$ of the system. We consider hydraulic
fractures at depth such that the confining stress is assumed to be
sufficiently large for the presence of a fluid lag to be negligible
(see discussion in \citet{GaDe00,LeDe07,Deto16}). In this limit,
the boundary conditions at the fracture front reduce to a zero fluid
flux normal to the front ${\displaystyle \left(\mathbf{q}\left(x_{c},z_{c}\right)=0\right)}$
and zero fracture width ${\displaystyle \left(w\left(x_{c},z_{c}\right)=0\right)}$
(see \citet{DePe14} for a detailed discussion).

Finally, the fracture is assumed to propagate in quasi-static equilibrium
under the assumption of linear elastic fracture mechanics (LEFM).
For a pure opening mode fracture, the propagation criterion reduces
to
\begin{equation}
\left(K_{I}\left(x_{c},z_{c}\right)-K_{Ic}\right)v_{c}\left(x_{c},z_{c}\right)=0\qquad v_{c}\left(x_{c},z_{c}\right)\ge0\qquad K_{I}\left(x_{c},z_{c}\right)\le K_{Ic}\label{eq:S2:PropCond}
\end{equation}
for all ${\displaystyle \left(x_{c},z_{c}\right)\in{\displaystyle \mathcal{C}\left(t\right)}}$.
In this equation, ${\displaystyle K_{I}}$ is the stress intensity
factor, ${\displaystyle K_{Ic}}$ the material fracture toughness,
and ${\displaystyle v_{c}\left(x_{c},z_{c}\right)}$ the local fracture
velocity normal to the front (see figure \ref{fig:Sketch}). When
the fracture is propagating at a point ${\displaystyle \left(x_{c},z_{c}\right)}$,
the velocity is positive, and the stress intensity factor equals the
material toughness (${\displaystyle v_{c}\left(x_{c},z_{c}\right)>0}$,
$K_{I}\left(x_{c},z_{c}\right)=K_{Ic}$).

\subsection{Numerical solver\label{subsec:S2:NumSolv}}

For the numerical solution of the moving boundary problem presented
in section \ref{subsec:S2:MathForm}, we use the open-source 3D-planar
hydraulic fracture solver PyFrac \citep{ZiLe20}. This solver is based
on the implicit level set algorithm (ILSA) originally developed by
\citet{PeDe08} for three-dimensional planar hydraulic fractures (see
also \citet{DoPe17} for more details). The numerical scheme combines
the discretization of a finite domain with the steadily moving plane-strain
hydraulic fracture asymptotic solution \citep{GaDe11} near the fracture
front. Even with a coarse discretization of the finite domain, the coupling between
these two scales allows for an accurate estimation of the fracture
front velocity $v_{c}\left(x_{c},z_{c}\right)$. We use the improvement
of \citet{PeLe21}, which imposes strict continuity of the fracture
front during its reconstruction from the level set values at the cell
center. The discretization of the elasticity equation (\ref{eq:S2:Elasticity})
is performed using piece-wise constant rectangular displacement discontinuity
elements, while an implicit finite volume scheme is used for elastohydrodynamic
lubrication flow. In various implementations, this numerical scheme
has proved to be both accurate and robust when tested against known
hydraulic fracture growth solutions \citep{Peir15,Peir16,ZiLe18,ZiLe20,MoLe20,MoLe21}.

We use a minimal initial discretization of $61\textrm{x}61$ elements
and add elements as the fracture elongates for all simulations presented
herein. Our simulations need to run over several orders of magnitude
in time and space to capture the transition and the late-time buoyant
propagation stage. We thus adopt two different remeshing techniques
to ensure that the smaller spatial dimension (horizontal in our case)
always satisfies a minimum discretization of ${\displaystyle 61}$
elements. A second condition of the discretization is that the original
element-aspect-ratio is ensured during the entire simulation, even
when the aspect ratio of the mesh domain is changing. This discretization
constrains the maximum relative error on the fracture radius to $2-3\%$
for radial fractures \citep{ZiLe20,MoLe21}. The fracture is initialized
as a radial hydraulic fracture in the viscosity-dominated regime \citep{SaDe02},
which corresponds to the early time solution of this type of fracture.
We use this technique to ensure that we consistently capture the entire
propagation in all the different regimes. 

\subsection{Scaling analysis}

In the configuration studied herein, the hydraulic fracture initially
propagates radially outwards from a point source. It remains radial
as long as the fracture is sufficiently small that buoyancy forces
remain negligible. At late time, the fracture elongates in the direction
of the buoyant force. A head and tail structure similar to the plane-strain
(2D) case is expected to develop. This head-tail structure has either a horizontal breadth
which stabilizes in space at late times or an ever-growing one %albeit with a possibly finite or ever-growing horizontal breadth 
 \citep{List90b,GaGe14}. We capture
the evolution of the fracture shape by introducing ${\displaystyle \ell\left(t\right)}$
as the vertical extent (to which we will alternatively refer as the
fracture length) and ${\displaystyle b\left(z,t\right)}$ as the horizontal
breadth (see figure \ref{fig:Sketch}). We recognize that the horizontal
breadth may not be uniform in space and will thus refer to ${\displaystyle b\left(t\right)}$
as the maximum horizontal breadth of the fracture. We scale these
fracture dimensions as
\begin{equation}
\ell(t)=\ell_{*}(t)\gamma(\mathcal{P}_{i}),\hspace{1em}b(t)=b_{*}(t)\beta(\mathcal{\mathcal{P}}_{i})\label{eq:S2:ScalingLandB}
\end{equation}
where ${\displaystyle \ell_{*}\left(t\right)}$ and ${\displaystyle b_{*}\left(t\right)}$
are a characteristic fracture length and (maximum) breadth respectively,
and ${\displaystyle \gamma}$, and ${\displaystyle \beta}$ the corresponding
dimensionless extent. Following the notation of previous contributions \citep{Deto04},
we scale the fracture width and net pressure as 
\begin{equation}
w\left(x,z,t\right)=w_{*}\left(t\right)\Omega\left(x/b_{*},z/\ell_{*},\mathcal{\mathcal{P}}_{i}\right)\qquad p\left(x,z,t\right)=p_{*}\left(t\right)\Pi\left(x/b_{*},z/\ell_{*},\mathcal{\mathcal{P}}_{i}\right)\label{eq:S2:ScalingWandP}
\end{equation}
with ${\displaystyle w_{*}\left(t\right)}$ and ${\displaystyle p_{*}\left(t\right)}$
the characteristic width and net pressure scales, ${\displaystyle \Omega}$
and ${\displaystyle \Pi}$ are the dimensionless width and pressure.
In the previous expressions, we recognized that the characteristic
scales may depend on time and that the dimensionless solution is a
function of a finite set of dimensionless numbers $\mathcal{P}_{i}$.

Introducing such a scaling into the governing equations provides a
set of dimensionless groups denoted by ${\displaystyle \mathcal{G}}$.
In particular, the scaling of the elasticity equation (\ref{eq:S2:Elasticity})
provides, besides the characteristic aspect ratio of the fracture
\[
{\displaystyle \mathcal{G}_{s}=b_{*}/\ell_{*}},
\]
a dimensionless group defined as the ratio between the characteristic
elastic pressure $w_{*}E^{\prime}/b^{*}$ and the characteristic net
pressure $p_{*}$
\begin{equation}
\mathcal{G}_{e}=\frac{w_{*}E^{\prime}}{p_{*}b_{*}}.\label{eq:S2:DefinitionGe}
\end{equation}
Elasticity is always of first order for a fracture problem (i.e. $\mathcal{G}_{e}=1$),
such that this equation yields a direct relation between the characteristic
net pressure, fracture opening, and a fracture dimension. Scaling
wise, the global volume conservation (\ref{eq:2SBala}) provides a
ratio between the released volume ${\displaystyle Q_{o}t}$ and the
characteristic fracture volume ${\displaystyle w_{*}b_{*}\ell_{*}}$
\begin{equation}
\mathcal{G}_{v}=\frac{Q_{o}t}{w_{*}b_{*}\ell_{*}}.\label{eq:S2:DefinitionGv}
\end{equation}
A dimensionless fracture toughness ${\displaystyle \mathcal{G}_{k}}$
emerges from the linear fracture propagation criteria $K_{I}=K_{Ic}$
as a ratio between the characteristic linear elastic fracture mechanics
pressure for the material ${\displaystyle K_{Ic}/\sqrt{b_{*}}}$ and
the characteristic net pressure ${\displaystyle p_{*}}$
\begin{equation}
\mathcal{G}_{k}=\frac{K_{Ic}}{p_{*}\sqrt{b_{*}}}.\label{eq:S2:DefinitionGk}
\end{equation}
Poiseuille's viscous drop (\ref{eq:S2:FluidFlow}) inside the fracture
provides a dimensionless group akin to a dimensionless viscosity defined
as the ratio between the characteristic viscous pressure ${\displaystyle \mu^{\prime}Q_{o}/w_{*}^{3}}$
and the characteristic pressure ${\displaystyle p_{*}}$ 
\begin{equation}
\mathcal{G}_{m}=\frac{\mu^{\prime}Q_{o}}{w_{*}^{3}p_{*}}.\label{eq:S2:DefinitionGm}
\end{equation}
Finally, a last dimensionless group relates the characteristic buoyancy
pressure ${\displaystyle \Delta\gamma\ell_{*}}$ to the characteristic
pressure $p_{*}$
\begin{equation}
\mathcal{G}_{b}=\frac{\Delta\gamma\ell_{*}}{p_{*}}.\label{eq:S2:DefinitionGb}
\end{equation}
Using these dimensionless groups to emphasize the relative importance
of the underlying physical mechanism, one obtains different scalings
associated with different propagation regimes.

\section{Onset of the buoyant regime\label{sec:OnsetOfBuoyant}}
The contribution of buoyant forces is negligible for a small enough
fracture: from (\ref{eq:S2:DefinitionGb}), \textbf{$\mathcal{G}_{b}\ll1$}.
In the absence of buoyancy, the HF propagates
with a radial penny-shaped geometry. In an impermeable medium, \citet{SaDe02}
have shown that the HF transitions from a viscosity-dominated regime
at early time towards a toughness-dominated regime at late time. The
increase in fracture energy dissipation is directly related to the
increase of the fracture perimeter. Self-similar solutions have been
obtained in both the ${\displaystyle \textrm{M}}$/viscous scaling
and the ${\displaystyle \textrm{K}}$/toughness scaling. Following
\citet{SaDe02}, the characteristic scales are denoted with a subscript
${\displaystyle m}$ for the ${\displaystyle \textrm{M}}$/viscous
scaling, and ${\displaystyle k}$ for the ${\displaystyle \textrm{K}}$/toughness
scaling (see table \ref{Apptab:RadialScales} in
appendix \ref{Appsec:ScalingsTables}). The transition from the early
time viscosity dominated to the toughness dominated regime is entirely
captured by a dimensionless toughness ${\displaystyle \mathcal{K}_{m}}$
increasing with time as \citep{SaDe02}
\begin{equation}
{\displaystyle \mathcal{K}_{m}=K_{Ic}\frac{t^{1/9}}{E^{\prime13/18}Q_{o}^{1/6}\mu^{\prime5/18}}}.\label{eq:S3:RadialKm}
\end{equation}
This dimensionless toughness (defined in the ${\displaystyle \textrm{M}}$-scaling)
is directly related to a dimensionless viscosity defined in the ${\displaystyle \textrm{K}}$-scaling
\begin{align}
\mathcal{M}_{k} & =\mathcal{K}_{m}^{-18/5}=\left(t_{mk}/t\right)^{2/5}.\label{eq:S3:RadialMk}
\end{align}
In the absence of buoyancy, the toughness-dominated regime is reached
when ${\displaystyle \mathcal{K}_{m}\sim\mathcal{M}_{k}\sim1}$ \citep{SaDe02}
(note our use of the fracture toughness $K_{Ic}$ instead of the reduced
fracture toughness used in some previous work $K^{\prime}=\sqrt{32/\pi}K_{Ic}$),
or alternatively for times greater than a characteristic time ${\displaystyle t_{mk}}$
defined as the time when ${\displaystyle \mathcal{K}_{m}=\mathcal{M}_{k}=1}$
\begin{equation}
t_{mk}=\frac{E^{\prime13/2}\mu^{\prime5/2}Q_{o}^{3/2}}{K_{Ic}^{9}}.\label{eq:S3:Definitiontmk}
\end{equation}
The corresponding characteristic fracture radius at this time of transition
between viscous and toughness growth is, according to \citet{SaDe02}
\begin{equation}
\ell_{mk}=\frac{E^{\prime3}Q_{o}\mu^{\prime}}{K_{Ic}^{4}}.\label{eq:S3:Definitionlmk}
\end{equation}
To estimate when the buoyancy forces will start to play a role, still
assuming that ${\displaystyle b_{*}\sim\ell_{*}}$ - a hypothesis
valid at the onset of the buoyant regime -, it is worth computing
the dimensionless buoyancy $\mathcal{G}_{b}$ (\ref{eq:S2:DefinitionGb}):
\begin{equation}
{\displaystyle \mathcal{B}_{m}=\varDelta\gamma\frac{Q_{o}^{1/3}t^{7/9}}{E^{\prime5/9}\mu^{\prime4/9}},\qquad\mathcal{B}_{k}=\varDelta\gamma\frac{E^{\prime3/5}Q_{o}^{3/5}t^{3/5}}{K_{Ic}^{8/5}}}\label{eq:S3:DefinitionBmAndBk}
\end{equation}
in the viscous (subscript $m$) and toughness (subscript $k$) scaling
respectively. As expected, the effect of buoyancy increases with time
as the fracture grows. For each limiting regime, we deduce a transition
time-scale where buoyancy becomes dominant as the time when $\mathcal{B}_{m}$
(respectively $\mathcal{B}_{k}$) equals one: 
\begin{equation}
{\displaystyle t_{m\widehat{m}}=\frac{E^{\prime5/7}\mu^{\prime4/7}}{\varDelta\gamma^{9/7}Q_{o}^{3/7}},\qquad t_{k\widehat{k}}=\frac{K_{Ic}^{8/3}}{E^{\prime}Q_{o}\varDelta\gamma^{5/3}}}.\label{eq:S3:DefinitionTmmAndTkk}
\end{equation}
In the following, we use a ${\displaystyle \widehat{\cdot}}$ to highlight
scalings where buoyancy plays a dominant role. Similarly to the previous
viscosity to toughness transition, it is practical to obtain the corresponding
transition length scales (see table \ref{Apptab:TransitionScales}
in appendix \ref{Appsec:ScalingsTables} for details)
\begin{equation}
{\displaystyle \ell_{m\widehat{m}}=\frac{E^{\prime3/7}Q_{o}^{1/7}\mu^{\prime1/7}}{\varDelta\gamma^{4/7}},\qquad\ell_{k\widehat{k}}=\frac{K_{Ic}^{2/3}}{\varDelta\gamma^{2/3}}\equiv\ell_{b}.}\label{eq:S3:DefinitionLmmAndLkk}
\end{equation}
It is worth noting that the toughness-buoyancy length scale ${\displaystyle \ell_{k\widehat{k}}}$
- that we will alternatively refer to as $\ell_{b}$ - can be directly
obtained by assuming ${\displaystyle b_{*}\sim\ell_{*}}$ and balancing
the toughness pressure ${\displaystyle K_{Ic}/\sqrt{\ell_{*}}}$ with
the buoyancy pressure ${\displaystyle \Delta\gamma\ell_{*}}$. Such
a buoyancy length scale $\ell_{b}$ is strictly equal to the one obtained
in the 2D plane-strain case \citep{Weer71,List90,LiKe91,RoLi07,HeOl94}
as well as for a finger-like three-dimensional geometry \citep{GaGe14}. 

The buoyancy effect becomes of order one either when the initially radial hydraulic fracture is still propagating in the viscous (which implies $\mathcal{K}_m(t=t_{m\widehat{m}}) < 1$(\ref{eq:S3:RadialKm})) or when it is already in the toughness-dominated regime (for which $\mathcal{M}_k(t=t_{k\widehat{k}}) < 1$ (\ref{eq:S3:RadialMk})). The interplay between the radial transition from viscosity- to toughness-dominated and the one from radial to buoyant can thus be captured by either
\begin{equation}
\mathcal{K}_{\widehat{m}}=\mathcal{K}_m\left(t = t_{m\widehat{m}}\right)=  \frac{K_{Ic}}{E^{\prime 9/14} Q_o^{3/14}\varDelta\gamma^{1/7}\mu^{\prime 3/14}} =  \left(\frac{\ell_{m\widehat{m}}}{ \ell_{mk}} \right)^{1/4} = \left(\frac{t_{m\widehat{m}}}{t_{mk}} \right)^{1/9}
\label{eq:S3:DefinitionKkhat}
\end{equation}
or 
\begin{equation}
\mathcal{M}_{\widehat{k}}=\mathcal{M}_k\left(t = t_{k\widehat{k}}\right)=\mu^{\prime}\frac{Q_{o}E^{\prime3}\varDelta\gamma^{2/3}}{K_{Ic}^{14/3}}=\frac{\ell_{mk}}{\ell_{k\hat{k}}} = \left(\frac{t_{mk}}{t_{k\widehat{k}}} \right)^{2/5}.
\label{eq:S3:DefinitionMkhat}
\end{equation}
These two dimensionless numbers are related as $\mathcal{M}_{\widehat{k}}^{-3/14}=\mathcal{K}_{\widehat{m}}$. In fact, the different transition time-scales(\ref{eq:S3:DefinitionTmmAndTkk}), and (\ref{eq:S3:Definitiontmk}) are related as $t_{m\widehat{m}}/t_{mk} =(t_{k\widehat{k}}/t_{mk})^{27/35} $. The transition to buoyancy can therefore be grasped by any ratio of these transition time-scales, such that only one of the two parameters of equations (\ref{eq:S3:DefinitionKkhat}) and (\ref{eq:S3:DefinitionMkhat}) is required to define the transition.

In the following, we choose $\mathcal{M}_{\widehat{k}}$ to quantify the transition from a radial to a buoyant hydraulic fracture.
Physically, $\mathcal{M}_{\widehat{k}}$ quantifies if the fracture is
viscosity- (${\displaystyle \mathcal{M}_{\widehat{k}}>1}$) or toughness-dominated
(${\displaystyle \mathcal{M}_{\widehat{k}}<1}$) at the onset of the
buoyant regimes. Interestingly, $\mathcal{M}_{\widehat{k}}$ is directly
the ratio of the characteristic viscous-toughness transition length
scale $\ell_{mk}$ (without buoyancy) with the buoyant toughness transition
scale $\ell_{b}=\ell_{k\widehat{k}}$. This confirms that $\mathcal{M}_{\widehat{k}}$
(\ref{eq:S3:DefinitionMkhat}) properly captures the competition between
the transition from viscous to toughness growth and the transition
to the buoyant regime. 

\section{Toughness-dominated buoyant fractures $\mathcal{M}_{\widehat{k}}\ll1$
\label{sec:ToughnessDominated}}

We first focus on toughness-dominated buoyant fractures ($\mathcal{M}_{\widehat{k}}\ll1$),
for which the transition to the buoyant regime occurs when the initially
radial fracture is already propagating in the toughness dominated
regime ($t_{k\widehat{k}}\gg t_{mk}$). Figures \ref{fig:ToughFp}e-i
show the complete fracture evolution for a value of ${\displaystyle \mathcal{M}_{\hat{k}}\approx1.0\times10^{-3}}$.
The fracture is initially radial (figure \ref{fig:ToughFp}e), elongates
as buoyancy commences to act (figures \ref{fig:ToughFp}f and g), and ends-up
being akin to a finger-like fracture (figures \ref{fig:ToughFp}h
and i). It is worth noting that for $t>t_{k\widehat{k}}$, the breadth
is uniform such that the creation of new fracture surfaces only occurs
in the head region. This buoyant fracture exhibits a head-tail structure
qualitatively similar to the plane-strain 2D case \citep{List90,RoLi07}.
In the tail, the breadth is constant, and no new fracture surfaces
are created in the horizontal direction. This can be clearly observed
from Figure \ref{fig:ToughFp} (footprints i-h and the evolution of
the breadth). In other words, the head is toughness-dominated, while
in the tail only a viscous vertical flow is dissipating energy.
\begin{figure}
\centerline{\includegraphics[height=0.8\textheight]{Fig2_ToughFigureComplete.pdf}}
\caption{Toughness-dominated buoyant fracture. Green dashed lines in all figures
indicate the 3D $\widehat{\text{K}}$ GG, (2014) solution. a) Opening
along the centerline ${ w\left(0,z,t\right)/w_{\widehat{k}}^{\textrm{head}}}$
for a simulation with ${ \mathcal{M}_{\widehat{k}}=1.\times10^{-2}}$.
b) Net pressure along the centerline ${ p\left(0,z,t\right)/p_{\widehat{k}}^{\textrm{head}}}$
for the same simulation. c) Fracture length ${ \ell\left(t\right)/\ell_{b}}$
for three simulations with large toughness ${ \mathcal{M}_{\widehat{k}}\in\left[10^{-3},10^{-1}\right]}$.
Dashed-dotted green lines highlight the late time linear term of the
${\displaystyle \widehat{\textrm{K}}}$ solution. d) Fracture breadth
${\displaystyle b\left(t\right)/\ell_{b}}$ (continuous) and head
breadth ${\displaystyle b^{\textrm{head}}\left(t\right)/\ell_{b}}$
(dashed). Grey lines  an error margin of ${\displaystyle 5\%}$. e
- i) Evolution of the fracture footprint from radial (e) towards the
final finger-like shape (h and i) for a fracture with ${\displaystyle \mathcal{M}_{\widehat{k}}=1.\times10^{-3}}$.
For the fracture shape in i), the vertical extent is cropped between
${\displaystyle \ell\left(t\right)/\ell_{b}=6}$ and ${\displaystyle \ell\left(t\right)/\ell_{b}=30}$.
Thick red dashed lines indicate the head shape according to the 3D $\widehat{\text{K}}$ GG, (2014) solution.
Note that the final stage i) has not reached the constant terminal
velocity (see inset c).\label{fig:ToughFp}}
\end{figure}

\textit{Toughness-dominated head} 

The characteristic scales of the toughness-dominated head are such
that $b_{*}^{\text{head}}\sim\ell_{*}^{\text{head}}$ and can be obtained
assuming that toughness, buoyancy, and elasticity are all of first-order
in the head. One obtains the following head scales:
\begin{align}
b_{\hat{k}}^{\text{head}} & =\ell_{\hat{k}}^{\text{head}}=\ell_{b}=\frac{K_{Ic}^{2/3}}{\varDelta\gamma^{2/3}},\hspace{1em}w_{\widehat{k}}^{\text{head}}=\frac{K_{Ic}^{4/3}}{E^{\prime}\Delta\gamma^{1/3}},\label{eq:S4:scalesToughHead}\\
p_{\hat{k}}^{\text{head}} & =K_{Ic}^{2/3}\Delta\gamma^{1/3},\hspace{1em}V_{\hat{k}}^{\textrm{head}}=Q_{o}t_{k\hat{k}}=\frac{K_{Ic}^{8/3}}{E^{\prime}\varDelta\gamma^{5/3}}.\nonumber 
\end{align}
which correspond exactly to the characteristic scales for a radial
hydraulic fracture at the transition to buoyancy $t=t_{k\hat{k}}.$
This scaling is similar (up to numerical factors) to those previously
obtained for 3D and 2D buoyant fractures \citep{List90,RoLi07,GaGe22}.

\textit{Viscosity-dominated tail}

The tail has a constant breadth equal to the characteristic breadth
scale of the head. In the tail, the viscous flow dissipation in the
vertical direction is quantified by the ratio of viscous pressure
${\displaystyle \mu^{\prime}v_{z*}\ell_{*}/w_{*}^{2}}$ to the characteristic
buoyancy pressure ${\displaystyle \varDelta\gamma\ell_{*}}$ (with
$\partial p/\partial z\ll\varDelta\gamma$ in the tail):
\begin{equation}
\mathcal{G}_{mz}=\frac{\mu^{\prime}v_{z*}}{w_{*}^{2}\varDelta\gamma},\label{eq:S2:DefinitionGmz}
\end{equation}
and is clearly dominant over any horizontal viscous dissipation. $\mathcal{G}{}_{mz}=1$
sets the characteristic vertical velocity as a function of the characteristic
tail opening. The elongated form of this buoyant fracture is such
that its aspect ratio is directly related to the ratio of characteristic
horizontal $v_{x*}$ to vertical $v_{z*}$ fluid velocities,
\begin{equation}
\frac{b_{*}}{\ell_{*}}\sim\frac{v_{x*}}{v_{z*}}.\label{eq:S4:DefinitionAspectRatio}
\end{equation}
and the characteristic vertical fracture velocity is of the same order
of magnitude as the vertical fluid velocity:
\begin{equation}
\frac{\partial\ell}{\partial t}\sim\frac{\ell_{*}}{t}=v_{z*}.\label{eq:S4:DefinitionGVelV}
\end{equation}
Assuming a viscosity-dominated tail of constant breadth $b_{*}=\ell_{b}$
set by buoyancy ($\mathcal{G}{}_{mz}=1$), global volume conservation,
elasticity ($\mathcal{G}_{v}=\mathcal{G}_{e}=1$), and equations (\ref{eq:S4:DefinitionAspectRatio})-(\ref{eq:S4:DefinitionGVelV})
provide the following characteristic tail scales:
\begin{align*}
\ell_{\widehat{k}}=\frac{Q_{o}^{2/3}\varDelta\gamma^{7/9}\,t}{K_{Ic}^{4/9}\mu^{\prime1/3}} & \qquad b_{\widehat{k}}=\ell_{b}\\
w_{\widehat{k}}=\frac{Q_{o}^{1/3}\mu^{\prime1/3}}{K_{Ic}^{2/9}\varDelta\gamma^{1/9}}=\mathcal{M}_{\widehat{k}}^{1/3} & w_{\widehat{k}}^{\text{head}}\qquad p_{\widehat{k}}=E^{\prime}\frac{\varDelta\gamma^{5/9}Q_{o}^{1/3}\mu^{\prime1/3}}{K_{Ic}^{8/9}}=\mathcal{M}_{\widehat{k}}^{1/3}p_{\widehat{k}}^{\text{head}}
\end{align*}
The corresponding horizontal characteristic fluid velocity decreases in inverse proportion to time as $v_{x*}=\ell_{b}/t$.

\subsection{Large time buoyant regime\label{subsec:S4:ToughnessLateTime}}

The head and tail structure of such a fracture with uniform breadth
can be further leveraged to obtain an approximate solution at late
time ($t\gg t_{k\widehat{k}}$) when assuming a state of plane strain
for each horizontal cross-section. Such an approximate 3D solution
was obtained by \citet{GaGe22}, imposing a toughness dominated head
and a viscosity dominated tail (in which $\partial p/\partial z\ll\varDelta\gamma$).
In that solution, that we will refer to as the 3D $\widehat{\text{K}}$ GG, (2014)
solution, the head is constant and the upward growth is governed by
the extension of the viscous tail. We compare numerical simulations
with this late time solution (this approximate solution in the scaling
used here is recalled in the supplementary material). We perform a
series of simulations for $\mathcal{M}_{\widehat{k}}=10^{-3},\,10^{-2}$
and $10^{-1}$. A typical evolution of the fracture opening and net
pressure along the centerline ${\displaystyle \left(x=0\right)}$
of a buoyant toughness fracture ($\mathcal{M}_{\widehat{k}}=10^{-2}$)
is reported in figures \ref{fig:ToughFp}a and b respectively. The
time evolution of length and breadth are illustrated in figure \ref{fig:ToughFp}c
and d. We can observe that both the fracture length and breadth compare
well with the 3D $\widehat{\text{K}}$ GG, (2014) solution at late time, especially
for $\mathcal{M}_{\widehat{k}}=10^{-3},\,10^{-2}$.

\begin{table}
\begin{center}
\def~{\hphantom{0}}
\begin{tabular}{ccccccccccccc}

${\displaystyle \mathcal{M}_{\widehat{k}}}$ & \multicolumn{4}{c}{${\displaystyle 10^{-3}}$} & \multicolumn{4}{c}{${\displaystyle 10^{-2}}$} & \multicolumn{4}{c}{${\displaystyle 10^{-1}}$}\\[3pt]

${\displaystyle t/t_{k\widehat{k}}}$ & ${\displaystyle 1.0}$ & ${\displaystyle 2.0}$ & ${\displaystyle 2.5}$ & ${\displaystyle 3.0}$ & ${\displaystyle 1.0}$ & ${\displaystyle 2.5}$ & ${\displaystyle 5.0}$ & ${\displaystyle 6}.0$ & ${\displaystyle 1.0}$ & ${\displaystyle 2.5}$ & ${\displaystyle 5.0}$ & $10.0$\\[2pt]

${\displaystyle \ell^{\textrm{head}}\left(t\right)/\ell_{b}}$ & ${\displaystyle 1.85}$ & ${\displaystyle 1.84}$ & ${\displaystyle 1.85}$ & ${\displaystyle 1.84}$ & ${\displaystyle 2.07}$ & ${\displaystyle 1.92}$ & ${\displaystyle 1.92}$ & ${\displaystyle 1.92}$ & ${\displaystyle 1.66}$ & ${\displaystyle 2.06}$ & ${\displaystyle 2.07}$ & ${\displaystyle 2.07}$\\[2pt]

mismatch with GG ($\%$) & ${\displaystyle 4.52}$ & ${\displaystyle 4.30}$ & ${\displaystyle 4.44}$ & ${\displaystyle 3.79}$ & ${\displaystyle 17.3}$ & ${\displaystyle 8.59}$ & ${\displaystyle 8.76}$ & ${\displaystyle 8.52}$ & ${\displaystyle 6.24}$ & ${\displaystyle 16.7}$ & ${\displaystyle 17.2}$ & ${\displaystyle 17.3}$\\[2pt]

${\displaystyle b^{\textrm{head}}\left(t\right)/\ell_{b}}$ & ${\displaystyle 0.68}$ & ${\displaystyle 0.68}$ & ${\displaystyle 0.68}$ & ${\displaystyle 0.68}$ & ${\displaystyle 0.72}$ & ${\displaystyle 0.72}$ & ${\displaystyle 0.72}$ & ${\displaystyle 0.72}$ & ${\displaystyle 0.78}$ & ${\displaystyle 0.84}$ & ${\displaystyle 0.84}$ & ${\displaystyle 0.84}$\\[2pt]

mismatch with GG ($\%$) & ${\displaystyle 0.52}$ & ${\displaystyle 0.60}$ & ${\displaystyle 0.56}$ & ${\displaystyle 0.54}$ & ${\displaystyle 5.19}$ & ${\displaystyle 5.32}$ & ${\displaystyle 5.29}$ & ${\displaystyle 5.36}$ & ${\displaystyle 13.9}$ & ${\displaystyle 23.1}$ & ${\displaystyle 23.3}$ & ${\displaystyle 23.3}$\\[2pt]

${\displaystyle V^{\textrm{head}}\left(t\right)/V_{\widehat{k}}^{\textrm{head}}}$ & ${\displaystyle 0.76}$ & ${\displaystyle 0.76}$ & ${\displaystyle 0.76}$ & ${\displaystyle 0.76}$ & ${\displaystyle 0.91}$ & ${\displaystyle 0.90}$ & ${\displaystyle 0.90}$ & ${\displaystyle 0.90}$ & ${\displaystyle 0.96}$ & ${\displaystyle 1.35}$ & ${\displaystyle 1.35}$ & ${\displaystyle 1.35}$\\[2pt]

mismatch withGG ($\%$) & ${\displaystyle 8.36}$ & ${\displaystyle 8.21}$ & ${\displaystyle 8.25}$ & ${\displaystyle 8.15}$ & ${\displaystyle 29.3}$ & ${\displaystyle 28.7}$ & ${\displaystyle 28.7}$ & ${\displaystyle 28.6}$ & ${\displaystyle 37.0}$ & ${\displaystyle 92.3}$ & ${\displaystyle 93.1}$ & ${\displaystyle 93.2}$\\[2pt]

$\ell^{\text{tail}}\left(t\right)/\ell_{b}$ & ${\displaystyle 3.60}$ & ${\displaystyle 17.5}$ & ${\displaystyle 24.4}$ & ${\displaystyle 31.3}$ & ${\displaystyle 0.89}$ & ${\displaystyle 10.2}$ & ${\displaystyle 25.7}$ & ${\displaystyle 31.9}$ & ${\displaystyle 0.271}$ & ${\displaystyle 3.43}$ & ${\displaystyle 9.86}$ & ${\displaystyle 22.8}$\\[2pt]

mismatch with GG($\%$) & ${\displaystyle 12.3}$ & ${\displaystyle 2.40}$ & ${\displaystyle 1.57}$ & ${\displaystyle 1.10}$ & ${\displaystyle 54.4}$ & ${\displaystyle 11.2}$ & ${\displaystyle 6.58}$ & ${\displaystyle 5.93}$ & ${\displaystyle 69.5}$ & ${\displaystyle 36.0}$ & ${\displaystyle 22.7}$ & ${\displaystyle 17.4}$\\
\end{tabular}
\caption{Comparison between characteristic head and tail length, head breadth
and head volume for toughness-dominated fractures ${  \mathcal{M}_{\widehat{k}}\in\left[10^{-3},10^{-1}\right]}$
at various dimensionless times ${  t/t_{k\widehat{k}}}$.
The mismatch is calculated as the relative difference between our
numerical results and the approximate 3D $\widehat{\text{K}}$ GG, (2014) solution
(GG in the table).\label{tab:ErrorsOnHeadTough}} 
\end{center}
\end{table}

\begin{figure}
\centering{\includegraphics[width=0.75\textwidth]{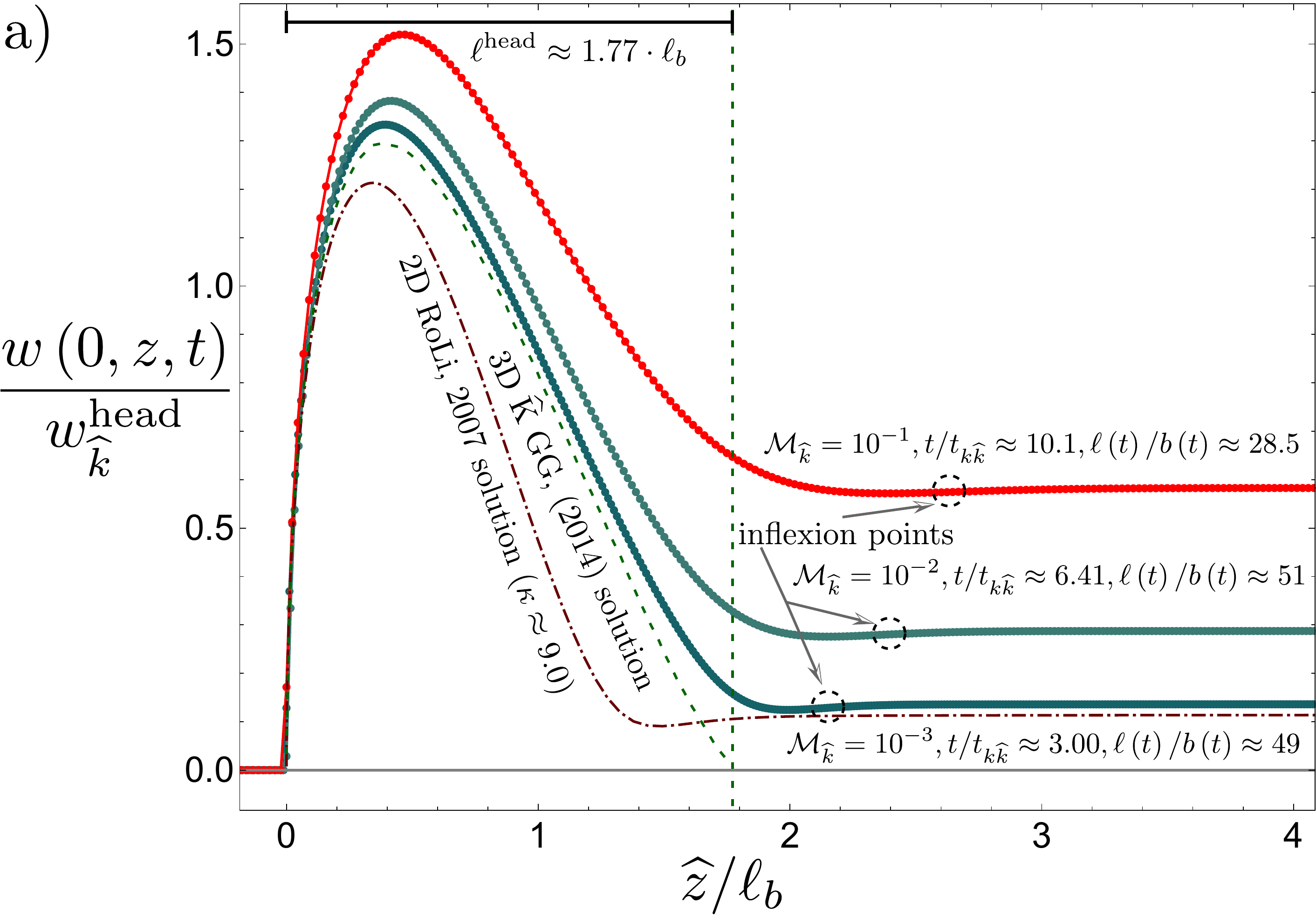}\ \includegraphics[width=0.75\textwidth]{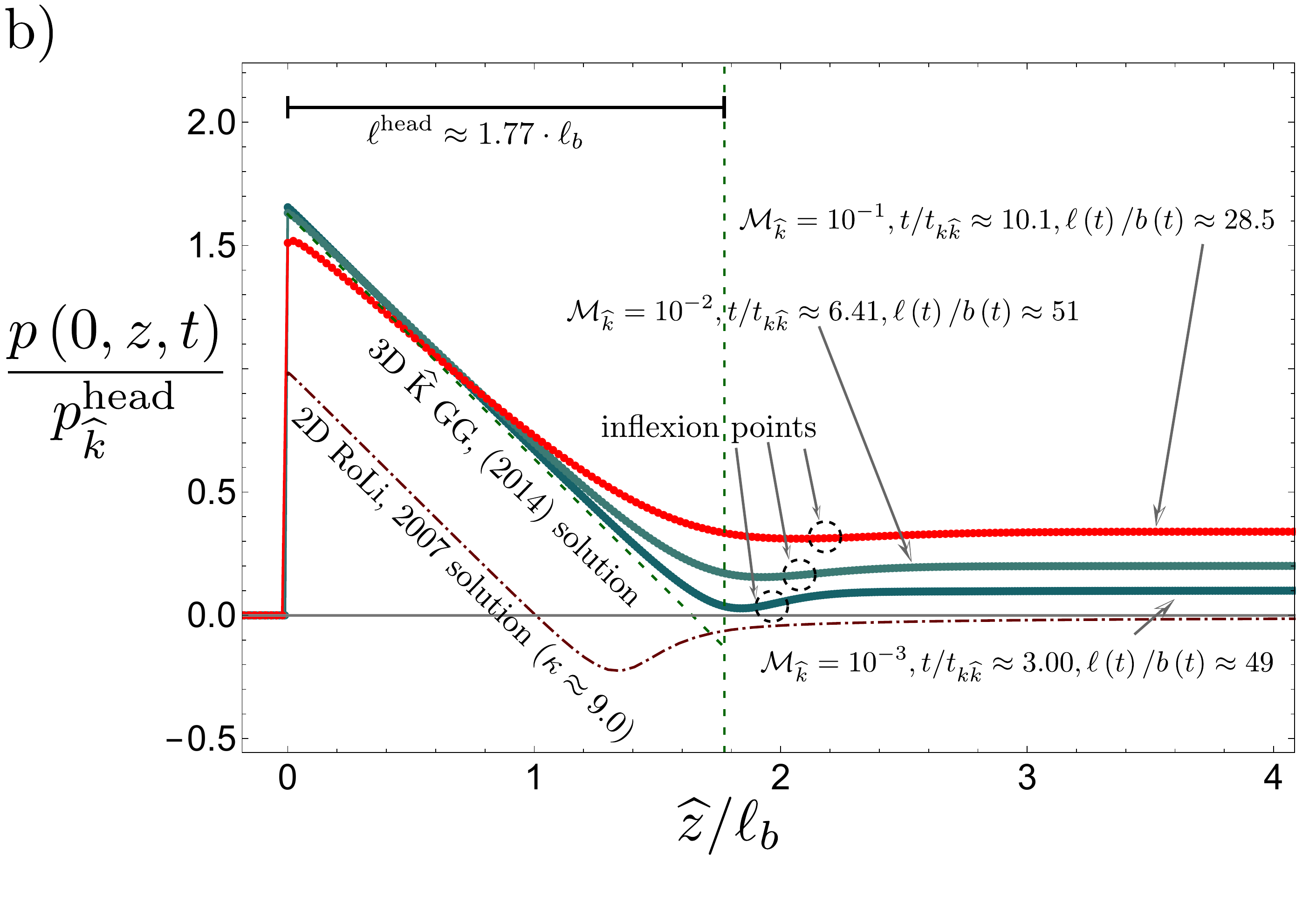} 
}
\caption{Tip based scaled opening (a) and pressure (b) of three toughness dominated
buoyant simulations with ${ \mathcal{M}_{\hat{k}}\in\left[10^{-3},10^{-1}\right]}$.
Continuous lines correspond to the PyFrac simulations \citep{ZiLe20}
with dots indicating the discretization (the number of elements in
the head is ${ >50}$), dashed lines to a 2D plane-strain
steadily moving solution. The vertical green dashed line indicates
the head length, and green continuous lines the 3D ${\displaystyle \widehat{\textrm{K}}}$
solutions.\label{fig:TipSolutionToughness}}
\end{figure}

We further compare various characteristic quantities from our simulations
with the 3D $\widehat{\text{K}}$ GG, (2014) late time solution of \citet{GaGe22}
in table \ref{tab:ErrorsOnHeadTough}. Our numerical evolution of
the head length ${\displaystyle \ell^{\textrm{head}}\left(t\right)/\ell_{b}}$
shows a marked variability but converges for the cases ${\displaystyle \mathcal{M}_{\widehat{k}}=1.\times10^{-3}}$
and ${\displaystyle \mathcal{M}_{\widehat{k}}=1.\times10^{-2}}$ to
their solution ${\displaystyle \ell^{\textrm{head}}\left(t\right)/\ell_{b}\sim1.77}$
at late time. The explanation for the variability lies within our
automatic evaluation of the head length from our numerical results.
Before an inflexion point forms in the opening along the centerline,
we estimate the head length as the maximum distance between the source
point and the front. Once an inflexion point forms (see figure \ref{fig:TipSolutionToughness}a),
we use either this inflexion point or a local pressure minimum between
the opening inflexion and the maximum pressure in the head (see figure
\ref{fig:TipSolutionToughness}b). These changes in criteria are more
visible for the less toughness-dominated simulation ${\displaystyle \mathcal{M}_{\widehat{k}}=1.\times10^{-1}}$.
Nonetheless, they do not affect the estimation of $\ell^{\textrm{head}}\left(t\right)$
for lower values of $\mathcal{M}_{\widehat{k}}$. Overall, the length
of the head stabilizes once it is evaluated via the pressure minima.
The reason is because \citet{GaGe22} similarly define the length
of the head as the point where the minimum pressure is reached (see
figure \ref{fig:TipSolutionToughness}). The relative difference of
${\displaystyle \sim4\%}$ for the simulation with ${\displaystyle \mathcal{M}_{\widehat{k}}=1.\times10^{-3}}$
is within the precision of our post-processing method. The increased
mismatch of ${\displaystyle \sim}8.5\%$ for ${\displaystyle \mathcal{M}_{\widehat{k}}=1.\times10^{-2}}$
is caused by a deviation from the strictly zero viscosity case and
the uncertainties of our evaluation method. Finally, the simulation
with ${\displaystyle \mathcal{M}_{\widehat{k}}=1.\times10^{-1}}$
has a relative difference ${\displaystyle \sim}17\%$, which clearly
reflects a significant deviation from the approximate 3D $\widehat{\text{K}}$ GG, (2014)
solution.

Defining the head breadth ${\displaystyle b^{\textrm{head}}=b(z=z^{\textrm{head}}=z_{Tip}-\ell^{\textrm{head}})}$
with ${\displaystyle z_{Tip}=\textrm{max}\left\{ z_{c}\right\} }$
(see figures \ref{fig:Sketch} and \ref{fig:ToughFp}i), figure \ref{fig:ToughFp}d
shows that the maximum breadth ${\displaystyle b\left(t\right)}$
(continuous lines) is equivalent to the head breadth ${\displaystyle b^{\textrm{head}}}$
(dashed lines) for ${\displaystyle \mathcal{M}_{\widehat{k}}\leq1.\times10^{-2}}$.
Combining these observations with figures \ref{fig:ToughFp}h and
i, we conclude that this breadth corresponds to the stabilized breadth
of the finger-like fracture. From figure \ref{fig:ToughFp}d, we observe
that the breadth in simulations ${\displaystyle \mathcal{M}_{\widehat{k}}=1.\times10^{-3}}$
and ${\displaystyle \mathcal{M}_{\widehat{k}}=1.\times10^{-2}}$ is
fully established for ${\displaystyle t/t_{k\hat{k}}\underset{\sim}{>}1}$,
corresponding to the moment where the head is entirely formed. This
is supported by the values displayed in table \ref{tab:ErrorsOnHeadTough} %{tab:ErrorsOnHeadTough}
that are stable for the corresponding simulations. We validate the
semi-analytical 3D $\widehat{\text{K}}$ GG, (2014) solution ${\displaystyle b\approx\pi^{-1/3}\ell_{b}}$
(green dotted line in figure \ref{fig:ToughFp}d) within our numerical
precision. The mismatch lies below ${\displaystyle <1\%}$ for ${\displaystyle \mathcal{M}_{\widehat{k}}=1.\times10^{-3}}$,
and is around ${\displaystyle \sim5\%}$ for ${\displaystyle \mathcal{M}_{\widehat{k}}=1.\times10^{-2}}$. 
For the simulation with ${\displaystyle \mathcal{M}_{\widehat{k}}=1.\times10^{-1}}$,
the breadth remains stable but shows a relative mismatch of about
${\displaystyle \sim25\%}$, indicating the limit of validity of the
3D $\widehat{\text{K}}$ GG, (2014) solution.

To ensure that the head is effectively constant in time, we additionally
estimate its volume. Generally, our estimated head volumes are larger
than the semi-analytical solution: ${\displaystyle V^{\textrm{head}}\approx0.701V_{\hat{k}}^{\textrm{head}}}$.
This phenomenon is not surprising as we overestimate the head length
with the post-processing of our numerical results. We can confirm
the emergence of a constant head volume and verify the order of magnitude
derived by \citet{GaGe22} for small values of ${\displaystyle \mathcal{M}_{\widehat{k}}}$
(\ref{eq:S3:DefinitionMkhat}). In conclusion, our numerical evaluation
indicates that the head of a buoyancy-driven hydraulic fracture is
constant and that the semi-analytical 3D $\widehat{\text{K}}$ GG, (2014) solution
of \citet{GaGe22} is valid as long as ${\displaystyle \mathcal{M}_{\widehat{k}}\leq1.\times10^{-2}}$.

It is interesting to compare the fully 3D results reported here with
the 2D plane-strain solutions previously reported in the literature
\citep{LiKe91,List90,RoLi07}. At late time, assuming that we are
far enough from the source region and neglecting any 3D curvature,
one can approximate the fracture as semi-infinite propagating at a
constant velocity. Such a two-dimensional solution has notably been
presented by \citet{RoLi07} for large toughnesses. Their scaling
can be retrieved from ours (\ref{eq:S4:scalesToughHead}) by replacing
the two-dimensional injection rate with ${\displaystyle Q_{2D}\sim\partial\ell_{\widehat{k}}/\partial t\,w_{\widehat{k}}}$.
We construct a two-dimensional numerical solver for a semi-infinite
hydraulic fracture combining a Gauss-Chebyshev quadrature for elasticity
and finite difference for lubrication flow similar to the one used
in \citet{MoLe18}. This 2D solver verifies exactly the large fracture
toughness limit of \citet{RoLi07}, and we use it to compare with
this contribution hereafter\textcolor{red}{{} }(we report details of
this 2D solver in the supplementary material).

In figure \ref{fig:TipSolutionToughness}, we plot the opening and
pressure along the centerline ${\displaystyle \left(x=0\right)}$
as a function of the tip based coordinate ${\displaystyle \widehat{z}\left(t\right)=z_{Tip}\left(t\right)-z}$,
such that ${\displaystyle \widehat{z}\left(t\right)\in\left[0,\ell\left(t\right)\right]}$
marks the interior of the fracture. Even for very small dimensionless
viscosities ${\displaystyle \left(\mathcal{M}_{\widehat{k}}\ll1\right)}$,
the pressure gradient in the head from the 3D numerical simulations
is not entirely linear and presents a gentler slope than the limiting 3D $\widehat{\text{K}}$ GG, (2014) solution (green dashed line \citep{GaGe22}).
Only for the simulation with ${\displaystyle \mathcal{M}_{\widehat{k}}=1.\times10^{-3}}$
is the viscous flow small enough to allow for a truly linear pressure
gradient in the head. The shape of the opening is qualitatively similar
between 2D and 3D (see ${\displaystyle \mathcal{M}_{\widehat{k}}=1.\times10^{-3}}$),
but the 2D ones shrink in the direction of the buoyant force. The
difference with the 2D solution is directly related to three-dimensional
effects associated with the curvature of the head.

The 3D \citet{GaGe22} and 2D \citet{RoLi07} solutions predict a
negative net pressure at the end of the head. Our 3D simulations do
not show such a feature and exhibit a smaller ``neck'' than the
one described by \citet{RoLi07} in 2D. The ``neck'' defines the
region at the end of the head, where fracture opening is reduced compared
to its stable value in the tail. This location is a pinch point
leading to the influx of the fluid from the tail into the head. Nevertheless,
figure \ref{fig:TipSolutionToughness} shows that the minimum pressure
in the neck decreases with a decreasing ${\displaystyle \mathcal{M}_{\widehat{k}}}$.
We expect that a negative net pressure should appear for smaller values
of ${\displaystyle \mathcal{M}_{\widehat{k}}}$. These observations
directly influence the opening distribution (figure \ref{fig:TipSolutionToughness}a).
We observe only a limited reduction of the opening between the tail
and the head in the fully 3D simulations. Nonetheless, such a neck
is present, and an inflexion point can be identified (black circles
in figure \ref{fig:TipSolutionToughness}a).  In the limit of zero
fluid viscosity, the opening in the tail would become ${\displaystyle 0}$.
This would be when the neck fully pinches and a finite volume pulse
forms.

\subsection{Transient toward the late buoyant regime \label{subsec:S4:TransitionToughness}}

In figure \ref{fig:ToughFp}c, an acceleration phase associated with
the transition to buoyancy can be observed. Such an acceleration is
directly related to the fact that, when radial, the fracture velocity
decreases with time as $\ell_{k}\propto t^{2/5}$ and ultimately,
once in the fully buoyant regime, reaches a constant velocity. The
intensity of such acceleration can be directly related to the dimensionless
number $\mathcal{M}_{\widehat{k}}$ by comparing this terminal velocity
with the radial velocity at the onset of buoyancy ${\displaystyle t=t_{k\widehat{k}}}$
(\ref{eq:S3:DefinitionTmmAndTkk}):
\begin{equation}
v_{\widehat{k}}/v_{k}(t_{k\widehat{k}})=\mathcal{M}_{\widehat{k}}^{-1/3}.\label{eq:S4:ToughnessOffset}
\end{equation}
The fracture needs to ``catch up'' from a length $\ell_{k}(t_{k\widehat{k}})\sim\ell_{b}$
to the buoyant late time solution ($\ell_{\widehat{k}}(t_{k\widehat{k}})\sim\mathcal{M}_{\widehat{k}}^{-1/3}\ell_{b}$)
and thus accelerates. According to figure \ref{fig:ToughFp}c, the
acceleration starts approximately when ${\displaystyle t/t_{k\widehat{k}}\approx0.5}$.
Correlating this with the observations of figure \ref{fig:ToughFp}a,
this corresponds approximately to the time when the bulk of the head
starts to leave the source region. The acceleration is thus driven
by the pressure difference between the head and tail visible in figure
\ref{fig:ToughFp}b. Figure \ref{fig:ToughFp}c further shows that
around ${\displaystyle t/t_{k\widehat{k}}\approx3}$, the fracture
starts to decelerate and approaches the complete 3D ${\displaystyle \widehat{\textrm{K}}}$
solution (green dashed lines). The simulation then presents a good
match until the end of the simulation (around ${\displaystyle t/t_{k\widehat{k}}\approx6.5}$).
A convergence towards the linear, dominant term (green dashed-dotted
lines) is only observed once a simulation reaches about ${\displaystyle t/t_{k\widehat{k}}\approx10}$
(see the simulation with ${\displaystyle \mathcal{M}_{\widehat{k}}=1.\times10^{-1}}$
in figure \ref{fig:ToughFp}c). This is consistent with the approximate
3D ${\displaystyle \widehat{\textrm{K}}}$ GG, (2014) solution which predicts
that the linear velocity is reached within ${\displaystyle 5\%}$
in relative terms when ${\displaystyle t/t_{k\widehat{k}}\approx14}$
(see supplementary material for details).

In the limiting case of zero fluid viscosity (${\displaystyle \mu^{\prime}=0\to\mathcal{M}_{\widehat{k}}=0}$),
the acceleration is infinite, and we can not hope to capture such
a sharp transition numerically. The strictly $\mathcal{M}_{\widehat{k}}=0$
limit corresponds to a three-dimensional Weertmans pulse \citep{Weer71}
associated with a zero-width tail. For very small but non-zero values
of ${\displaystyle \mu^{\prime}|\mathcal{M}_{\hat{k}}}$, overcoming
the transition phase is numerically challenging but possible. Defining
the end of the transient via the ${\displaystyle 5\%}$ deviation
level from the 3D approximate solution (${\displaystyle t/t_{k\widehat{k}}\approx14})$,
we obtain a corresponding fracture length of ${\displaystyle \ell\left(t\right)\sim19\mathcal{M}_{\widehat{k}}^{-1/3}\ell_{b}}$.
Expressing this limit as the aspect ratio ${\displaystyle \ell\left(t\right)/b\left(t\right)}$,
assuming that the breadth follows the \citet{GaGe22} solution (${\displaystyle b\left(t\right)\approx\pi^{-1/3}\ell_{b}})$,
the required aspect ratio is ${\displaystyle \ell\left(t\right)/b\left(t\right)\approx28\mathcal{M}_{\widehat{k}}^{-1/3}}$.
The numerical example with ${\displaystyle \mathcal{M}_{\widehat{k}}=1.\times10^{-2}}$
(largest value of ${\displaystyle \mathcal{M}_{\widehat{k}}}$ validating
the 3D ${\displaystyle \widehat{\textrm{K}}}$ solution) leads to
a aspect ratio of ${\displaystyle \ell\left(t\right)/b\left(t\right)\sim132}$
with a corresponding fracture length of ${\displaystyle \ell\left(t\right)\sim90\ell_{b}}$.
Such fracture lengths require a significant number of discretization
cells. Numerically, the discretization is mainly bounded by two parameters:
the distance of the source point to the fracture front and the number
of elements discretizing the head where a strong gradient of opening
and pressure takes place. In the toughness-dominated case the first
is more restrictive and requires discretizations of about ${\displaystyle 44}$
elements per ${\displaystyle \ell_{b}}$. The total number of degrees
of freedom thus quickly exceeds the current computational capacities
of PyFrac \citep{ZiLe20} and ultimately explains why we do not report
simulations for values of $\mathcal{M}_{\widehat{k}}$ lower than
$10^{-3}$.

\section{Viscosity-dominated buoyant fractures $\mathcal{M}_{\widehat{k}}\gg1$
\label{sec:ViscDominate}}

\begin{figure}
\centerline{\includegraphics[width=0.8\textwidth]{Fig4_ViscFigureComplete.pdf}}
\caption{Viscosity-dominated buoyant fracture. a) Opening along the centerline
${ w\left(x=0,z,t\right)/w_{m\widehat{m}}}$ for a simulation
with ${ \mathcal{M}_{\widehat{k}}=\infty}$. b) Net pressure
along the centerline ${ p\left(x=0,z,t\right)/p_{m\widehat{m}}}$
for the same simulation. c) Fracture length ${ \ell\left(t\right)/\ell_{m\widehat{m}}}$
for six simulations with large viscosity ${ \mathcal{M}_{\widehat{k}}\in\left[5.\times10^{2},\infty\right[}$.
d) Fracture breadth ${ b\left(t\right)/\ell_{m\widehat{m}}}$
for the same simulations. e - i) Evolution of the fracture footprint
from radial (e) towards the final elongated inverse cudgel shape (h
and i) for the same simulation as in a and b.\label{fig:ViscFp}}
\end{figure}
 We now turn to the viscosity dominated limit for which the transition
to buoyancy occurs prior to the transition to the radial toughness
dominated regime: $t_{m\widehat{m}}\ll t_{mk}$, i.e. $\mathcal{M}_{\widehat{k}}\gg1$.
We focus on the limiting case of a strictly zero-fracture toughness
(${\displaystyle \mathcal{M}_{\widehat{k}}=\infty}$), that we will
also refer to as the $\widehat{\text{M}}$ limit (at late time). The
evolution of such a fracture can be grasped from the numerical results
reported in figures \ref{fig:ViscFp}e-i. Similar to the toughness
case (figure \ref{fig:ToughFp}), the fracture is initially radial
(figure \ref{fig:ViscFp}e) and elongates (figures \ref{fig:ViscFp}f-i)
as soon as buoyancy plays a role ($t\sim t_{m\widehat{m}}$). The
overall footprint is strikingly different from the toughness limit.
Notably, the fracture breadth is not uniform along the vertical direction
and continuously grows horizontally due to the lack of any resistance
to fracture. The shape of the fracture at late time is akin to an
inverted cudgel with a distinct source and head regions.

\subsection{Late-time zero-toughness limit\label{subsec:ViscSS}}

It is enlightening to compare this simulation for $\mathcal{M}_{\widehat{k}}=\infty$
with the scaling originally derived by \citet{List90b} for this problem
(and his near-source solution). We first recall briefly the argument
of such a scaling. Contrary to the toughness limit, the breadth is
not constant, but the aspect ratio of the fracture remains related
to the ratio of the characteristic horizontal $v_{x*}$ to vertical
$v_{z*}$ fluid velocities
\begin{equation}
\frac{b_{*}}{\ell_{*}}\sim\frac{v_{x*}}{v_{z*}}.\label{eq:S4:DefinitionAspectRatio-1}
\end{equation}
The horizontal and vertical extent are linked to their corresponding
velocities as $b_{*}=v_{x*}t$, $\ell_{*}=v_{z*}t$. Viscous fluid
dissipation for viscous fractures occurs as much in the vertical as it does
in the horizontal direction. Vertically, the net pressure gradient
$\partial p/\partial z$ is negligible compared to $\varDelta\gamma$
such that, similarly to the viscosity dominated tail in the $\widehat{\text{K}}$
limit, the dimensionless ratio
\begin{equation}
\mathcal{G}_{mz}=\frac{\mu^{\prime}v_{z*}}{w_{*}^{2}\varDelta\gamma},\label{eq:S2:DefinitionGmz-1}
\end{equation}
is of order one. Horizontally, in the absence of gravitational forces,
the magnitude of viscous flow is quantified by the ratio of the horizontal
viscous pressure ${\displaystyle \mu^{\prime}v_{x*}b_{*}/w_{*}^{2}}$
to the characteristic net pressure ${\displaystyle p_{*}}$
\begin{equation}
\mathcal{G}_{mx}=\frac{\mu^{\prime}v_{x*}b_{*}}{w_{*}^{2}p_{*}}\label{eq:S2:DefinitionGmx}
\end{equation}
which is also of order one. Combined with elasticity ($\mathcal{G}_{e}=1$)
and global volume balance ($\mathcal{G}_{v}=1$), solving for the
lengths, width and pressure scales, we recover the scaling of \citet{List90b}:
\begin{align}
\ell_{\hat{m}}=\frac{\varDelta\gamma^{1/2}\,Q_{o}^{1/2}}{E^{\prime1/6}\mu^{\prime1/3}}t^{5/6},\qquad b_{\hat{m}}=\frac{E^{\prime1/4}Q_{o}^{1/4}}{\varDelta\gamma^{1/4}}t^{1/4},\label{eq:S5:DefinitionViscScales}\\
w_{\widehat{m}}=\frac{Q_{o}^{1/4}\mu^{\prime1/3}}{\varDelta\gamma^{1/4}\,E^{\prime1/12}}t^{-1/12},\qquad p_{\widehat{m}}=\frac{E^{\prime^{2/3}}\mu^{\prime1/3}}{t^{1/3}}.\nonumber 
\end{align}
Interestingly, in that scaling, the dimensionless toughness ($\mathcal{G}_{k}\equiv\mathcal{K}_{\widehat{m}}$)
associated with horizontal growth (defined with $b_{*}$ as the characteristic
fracture length) increases with time. From equation (\ref{eq:S2:DefinitionGk}),
we obtain the ``horizontal'' (subscript $x$) dimensionless toughness
\begin{equation}
\mathcal{K}_{\widehat{m},x}(t)=K_{Ic}\frac{\varDelta\gamma^{1/8}\,t^{5/24}}{E^{\prime19/24}Q_{o}^{1/8}\mu^{\prime1/3}}=\mathcal{M}_{\widehat{k}}^{-3/14}\left(\frac{t}{t_{m\widehat{m}}}\right)^{5/24}.\label{eq:S5:DefinitionKqx}
\end{equation}
As a result, for the case of finite fracture toughness, one expects
the horizontal growth to stop (and thus the breadth to stabilize)
when $\mathcal{K}_{\widehat{m},x}(t)$ reaches order one.
\begin{figure}
\centerline{\includegraphics[width=0.95\textwidth]{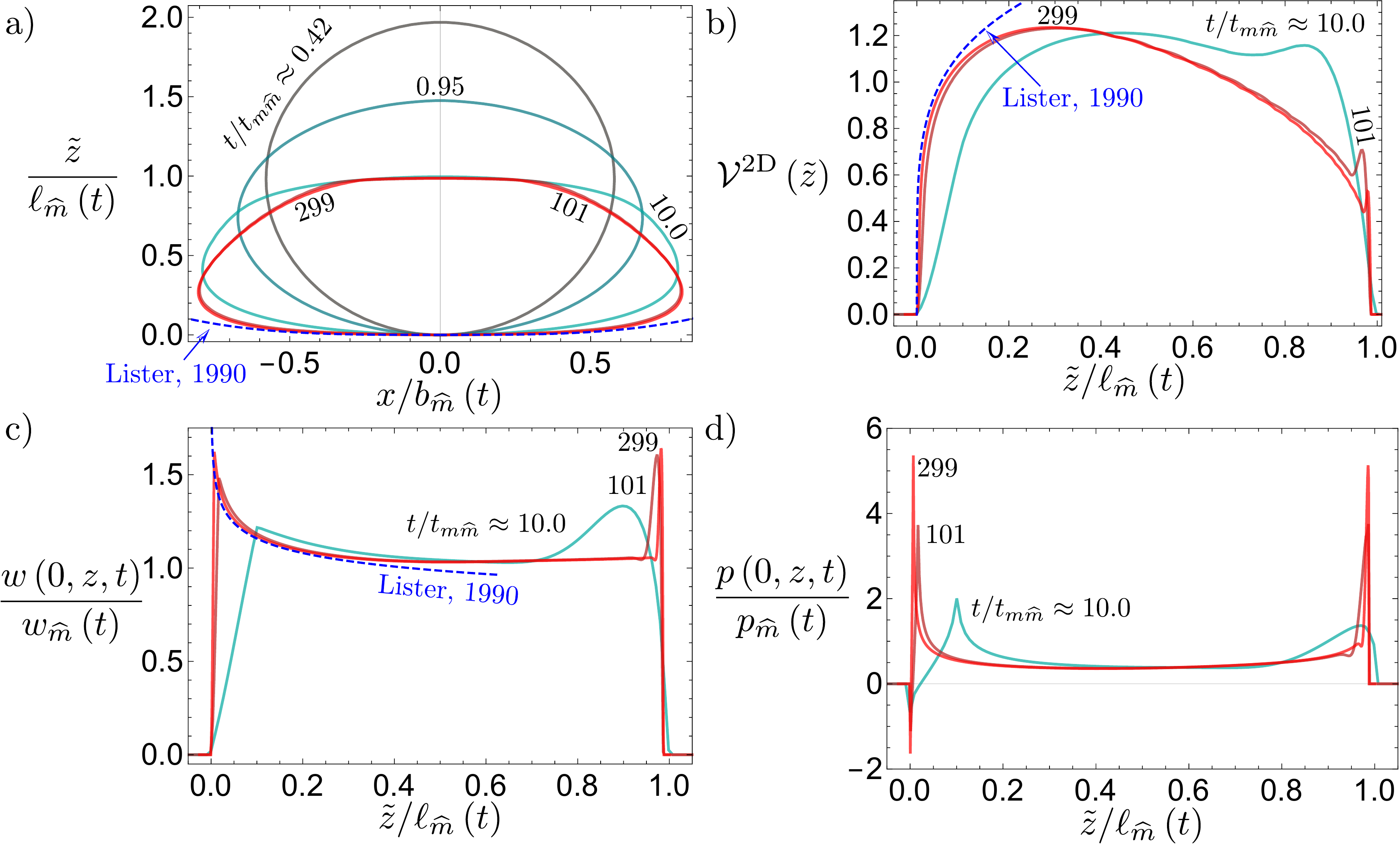}}
\caption{Scaled evolution of characteristic values of a buoancy-driven viscosity-dominated
fracture. Fracture footprint (a), cross-sectional volume (integral
of the opening over the breadth) (b), opening (c), and pressure (d)
at various dimensionless times ${\displaystyle t/t_{m\widehat{m}}}$
. Blue dashed lines represent the pseudo-three-dimensional near-source
solution of \citet{List90b}. A shifted coordinate system ${\displaystyle \tilde{z}}$
is used such that the lowest point of the fracture marks ${\displaystyle \tilde{z}=0}$.}
\label{fig:SelfSimVisc}
\end{figure}

\noindent The time evolution of fracture length and breadth obtained
numerically (figures \ref{fig:ViscFp}c and d) exhibit a transition
from the radial viscosity regime to this late buoyant viscous scaling.
The power-law evolution with time of length and breadth matches equation
(\ref{eq:S5:DefinitionViscScales}) precisely at late time for the
$\mathcal{M}_{\widehat{k}}=\infty$ simulation. Contrary to the toughness
case, where the horizontal growth stops abruptly, we observe a smoother
horizontal deceleration accompanied by vertical acceleration, which
is less abrupt than in the toughness case.

\begin{figure}
\centerline{\includegraphics[height=0.8\textheight]{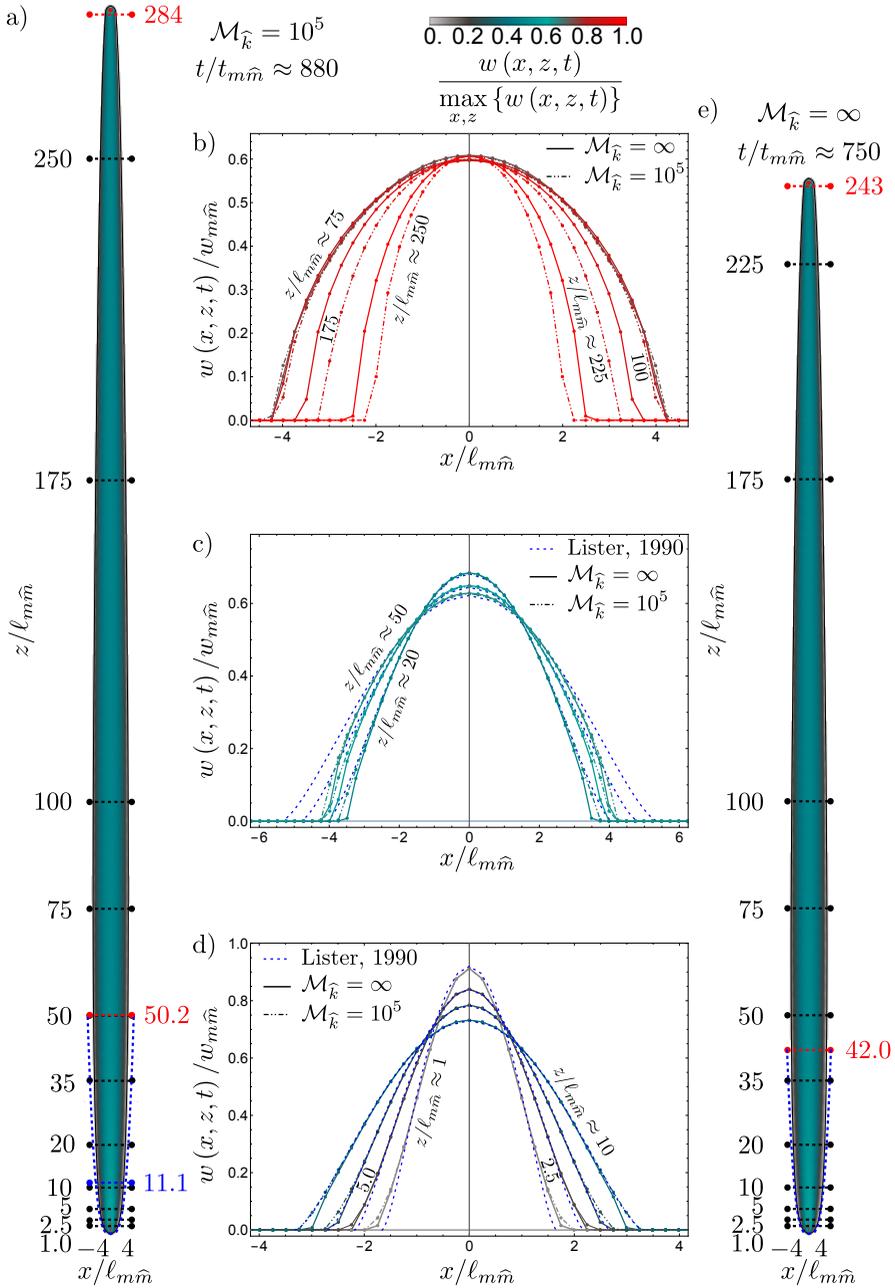}}
\caption{Footprint and cross-sectional opening profiles of two buoyant, viscosity-dominated
fractures. The colour code of the fractures represents the scaled
opening as described at the top. Black lines correspond to opening-profile
evaluations. The horizontal blue dashed line in a) is the limiting
height for the viscous solution of \citet{List90b}. Blue dashed lines
in a) and e) show the \citet{List90b} solution. Red dashed lines
mark the maximum breadth and the beginning of the head. Figures b
to d show the opening profiles in the cross-section where blue-dashed
lines represent the \citet{List90b} solution, dashed-dotted lines
correspond to ${  \mathcal{M}_{\widehat{k}}=\infty}$
and continuous lines to ${  \mathcal{M}_{\widehat{k}}=10^{5}}$.\label{fig:ListerFullFpComp}}
\end{figure}

In this zero toughness limit, at late time, the growth of the fracture
is self-similar and will not stop (neither horizontally nor vertically)
as long as the volume release continues. To confirm the overall self-similarity
of such a viscous, buoyant late time regime, we rescaled our numerical
results at different times and plot scaled footprints, centerline
width and net pressure, as well as the volume of each horizontal cross-section
in figure \ref{fig:SelfSimVisc}. The ${\displaystyle z}$-axis is
shifted such that the lowest point of the fracture coincides with
${\displaystyle \tilde{z}=0}$. A nice collapse of the scaled footprint
is observed for ${\displaystyle t/t_{m\widehat{m}}\underset{\sim}{>}100}$.
A similar collapse appears for centerline sections of width, pressure,
and cross-section volume. We recognize that the head region shrinks
with time and eventually reduces to a boundary layer. Before discussing
the head region, we observe that the source region solution derived
in \citet{List90b} matches our numerical results albeit in a relatively
narrow zone close to the injection point only. The \citet{List90b}
solution is based on a pseudo-three-dimensional approximation assuming
only horizontal growth with an unspecified upper ``head'' part.
In this approximate solution, the breadth increases monotonically
with the scaled coordinate $z/\ell_{\widehat{m}}(t)$ without any
possibility of reduction at large $z/\ell_{\widehat{m}}(t)$ to model
the fracture ``head''. For the \citet{List90b} solution, the distance
within which this source solution is applicable depends on the material,
fluid, and release properties. This distance is equivalent to the
transition length scale of a fracture without buoyancy $\ell_{mk}$,
which for the zero toughness case becomes infinite. This solution,
however, appears as the correct inner solution in the near-source
region (but not up to $z\sim\ell_{mk}$). Further comparison of the
width profiles at different cross-sections between our numerical solution
and this approximation is reported in figure \ref{fig:ListerFullFpComp}.

\noindent 

\subsubsection{Head region}

From both the footprints with width contours displayed in figure \ref{fig:ViscFp}
and the scaled profiles in figure \ref{fig:SelfSimVisc}, we observe
that, contrary to the toughness case, the head region shrinks with
time. Self-similarity of the overall fracture growth actually becomes
evident when the volumes of the head and the source region are negligible
compared to the volume in the tail, i.e. for times greater than $\sim100\,t_{m\widehat{m}}$.
The depletion of the head can be explored by the following scaling
argument. In a viscous head ($b_{*}^{\text{head}}\sim\ell_{*}^{\text{head}}$),
the horizontal and vertical fluid velocities are of the same order,
elasticity ($\mathcal{G}_{e}=1$), buoyancy ($\mathcal{G}_{b}=1$),
and viscous dissipation dominates ($\mathcal{G}_{mz}=1$), but its
volume is \textit{a priori} unknown. In addition, we assume that the
characteristic fluid velocity in the head is given by the vertical
characteristic velocity ${\displaystyle v_{z\widehat{m}}}\sim{\displaystyle \ell_{\widehat{m}}/t}$
from equation (\ref{eq:S5:DefinitionViscScales}). In other words,
the volumetric flow rate between the head and the tail is ${\displaystyle Q_{*}=w_{*}^{\text{head}}b_{*}^{\text{head}}v_{z\widehat{m}}}$.
Under those assumptions, the corresponding characteristic viscous
head scales are: 
\begin{align}
\ell_{\hat{m}}^{\text{head}}=b_{\hat{m}}^{\text{head}}=\frac{E^{\prime11/24}Q_{o}^{1/8}\mu^{\prime1/6}}{\varDelta\gamma^{5/8}t^{1/24}},\hspace{1em}w_{\hat{m}}^{\textrm{head}}=\frac{Q_{o}^{1/4}\mu^{\prime1/3}}{E^{\prime1/12}\varDelta\gamma^{1/4}t^{1/12}},\nonumber \\
p_{\widehat{m}}^{\textrm{head}}=\frac{E^{\prime11/24}Q_{o}^{1/8}\mu^{\prime1/6}\varDelta\gamma^{3/8}}{t^{1/24}},\hspace{1em}V_{\hat{m}}^{\textrm{head}}=\frac{E^{\prime5/6}Q_{o}^{1/2}\mu^{\prime2/3}}{\varDelta\gamma^{3/2}t^{1/6}}.\label{eq:S5:DefinitionViscHeadScales}
\end{align}
\begin{table}
\begin{center}
\def~{\hphantom{0}}
\begin{tabular}{ccccccccccccc}
${\displaystyle \mathcal{M}_{\widehat{k}}}$ & \multicolumn{4}{c}{${\displaystyle 10^{4}}$} & \multicolumn{4}{c}{${\displaystyle 10^{5}}$} & \multicolumn{4}{c}{${\displaystyle \infty}$}\\[3pt]

${\displaystyle t/t_{m\widehat{m}}}$ & ${\displaystyle 10}$ & ${\displaystyle 100}$ & ${\displaystyle 200}$ & ${\displaystyle 350}$ & ${\displaystyle 10}$ & ${\displaystyle 100}$ & ${\displaystyle 200}$ & ${\displaystyle 350}$ & ${\displaystyle 10}$ & ${\displaystyle 50}$ & ${\displaystyle 100}$ & ${\displaystyle 125}$\\[2pt]

${\displaystyle \ell^{\textrm{head}}\left(t\right)/\ell_{\hat{m}}^{\text{head}}\left(t\right)}$ & ${\displaystyle 2.39}$ & ${\displaystyle 2.33}$ & ${\displaystyle 2.29}$ & ${\displaystyle 2.33}$ & ${\displaystyle 2.39}$ & ${\displaystyle 2.37}$ & ${\displaystyle 2.31}$ & ${\displaystyle 2.21}$ & ${\displaystyle 2.63}$ & ${\displaystyle 2.86}$ & ${\displaystyle 2.79}$ & ${\displaystyle 2.77}$\\[2pt]

${\displaystyle V^{\textrm{head}}\left(t\right)/V_{\widehat{m}}^{\textrm{head}}\left(t\right)}$ & ${\displaystyle 4.66}$ & ${\displaystyle 5.34}$ & ${\displaystyle 5.38}$ & ${\displaystyle 5.58}$ & ${\displaystyle 4.66}$ & ${\displaystyle 5.34}$ & ${\displaystyle 5.29}$ & ${\displaystyle 5.08}$ & ${\displaystyle 5.07}$ & ${\displaystyle 5.78}$ & ${\displaystyle 5.63}$ & ${\displaystyle 5.56}$\\[2pt]

${\displaystyle w_{\textrm{max}}^{\textrm{head}}\left(t\right)/w_{\hat{m}}^{\text{head}}\left(t\right)}$ & ${\displaystyle 1.35}$ & ${\displaystyle 1.70}$ & ${\displaystyle 1.71}$ & ${\displaystyle 1.73}$ & ${\displaystyle 1.34}$ & ${\displaystyle 1.62}$ & ${\displaystyle 1.65}$ & ${\displaystyle 1.65}$ & ${\displaystyle 1.27}$ & ${\displaystyle 1.30}$ & ${\displaystyle 1.30}$ & ${\displaystyle 1.30}$\\
\end{tabular}
\caption{Comparison between characteristic head length, head volume, and maximum
opening in the head (${  w_{\textrm{max}}^{\textrm{head}}=\protect\underset{x,z}{\textrm{max}}\left\{ w\left(x,z\in\left[z_{tip}-\ell^{\textrm{head}}\left(t\right),z_{tip}\right],t\right)\right\} }$)
for viscosity-dominated fractures ${  \mathcal{M}_{\widehat{k}}\in\left[1\times10^{4},\infty\right[}$
at various dimensionless times ${  t/t_{m\widehat{m}}}$.\label{tab:ViscosityHead}}
\end{center}
\end{table}
These characteristic scales are consistent with the shrinking/depleting
viscous head observed numerically. The numerical validation is presented
in table \ref{tab:ViscosityHead}, where we observe the evolution
of the head length, head volume, and the maximum opening in the head.
Even though, we do not have an analytical or semi-analytical solution
to compare to, stabilization, when normalized with the depleting scales
\ref{eq:S5:DefinitionViscHeadScales}, is observed in table \ref{tab:ViscosityHead}
within the precision of our automatic evaluation of the head length.
It is interesting to note that at the onset of buoyancy, for ${\displaystyle t\approx t_{m\widehat{m}}}$
(defined in equation (\ref{eq:S3:DefinitionTmmAndTkk})), these scales
are strictly equal to the radial viscosity dominated scales (e.g.
$\ell_{\hat{m}}^{\text{head}}(t_{m\widehat{m}})=\ell_{m}(t_{m\widehat{m}})$,
$V_{\hat{m}}^{\textrm{head}}(t_{m\widehat{m}})=Q_{o}t_{m\widehat{m}}$).
This confirms the mechanism of a viscous head that detaches from the
source region and slowly depletes as it moves upward.

\textit{Comparison with the semi-infinite plane-strain solution} 

\begin{figure}
\centering{\includegraphics[width=0.8\textwidth]{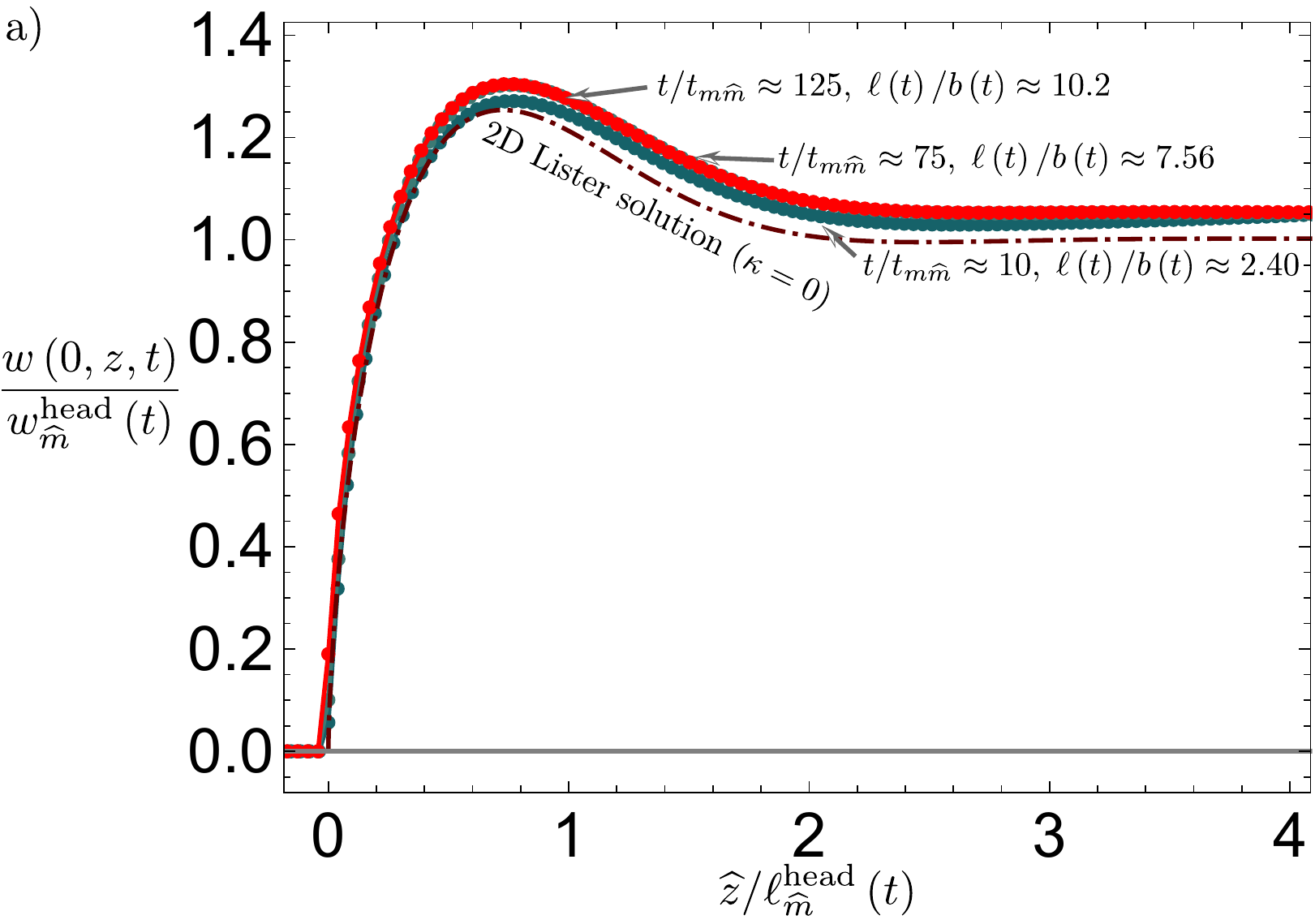}\smallskip{}
\includegraphics[width=0.8\textwidth]{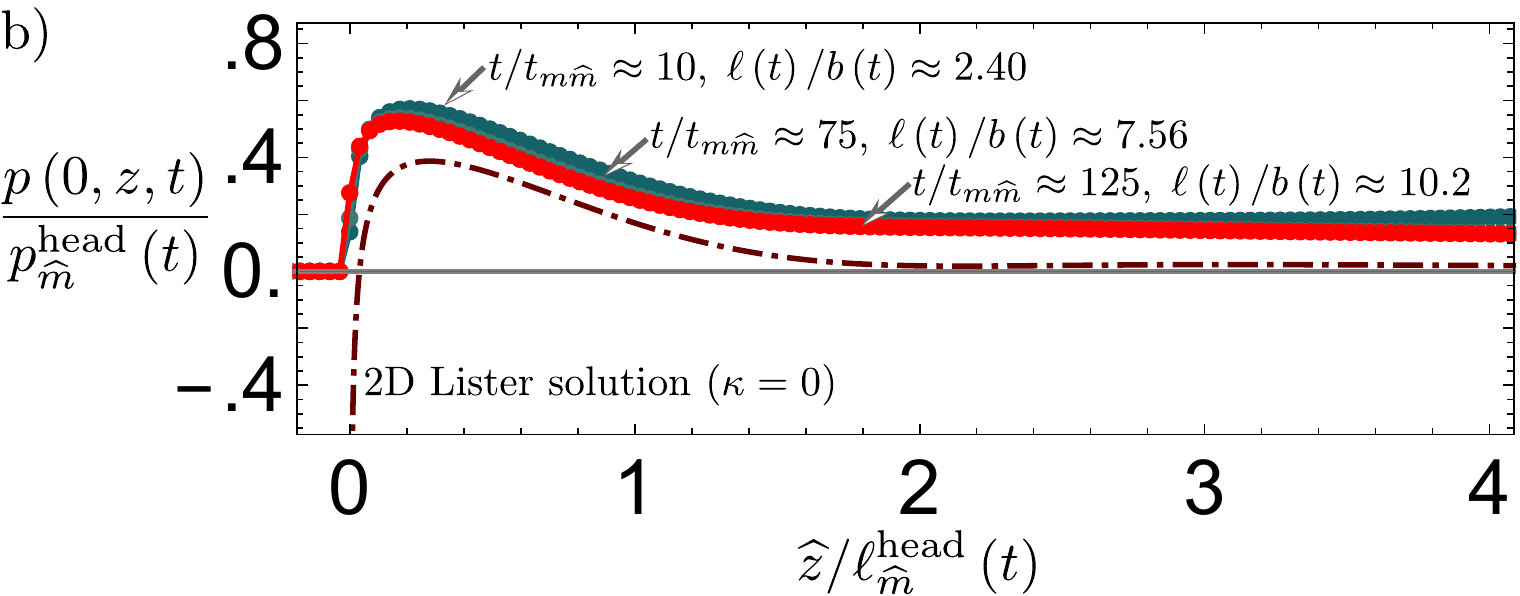}
}
\caption{Tip based opening (a) and pressure (b) of a viscosity-dominated buoyant
simulation with ${ \mathcal{M}_{\hat{k}}=\infty}$ in
function of the scaled tip coordinate. Continuous lines correspond
to the simulations with PyFrac \citep{ZiLe20} with dots marking the
location of discrete evaluations. The dotted-dashed line shows the
2D plane-strain steadily moving solution (see details in the supplementary
material ). \label{fig:TipBasedVisc}}
\end{figure}

Such a 3D viscous head can be compared to the existing 2D plane-strain
solution for a viscosity dominated steadily moving buoyant fracture
\citep{List90}. The 2D scales of \citet{List90} are based on a
constant fracture velocity. For the three-dimensional case, the characteristic
fracture velocity $v_{z\widehat{m}}$ decreases as
\[
v_{z\widehat{m}}=\frac{\ell_{\widehat{m}}}{t}=\frac{\varDelta\gamma^{1/2}\,Q_{o}^{1/2}}{E^{\prime1/6}\mu^{\prime1/3}t^{1/6}}
\]
which can be translated into a reducing two-dimensional release rate
by multiplication with the characteristic tail opening
\begin{equation}
Q_{2D}\sim v_{z\widehat{m}}w_{\widehat{m}}\sim\frac{Q_{o}^{3/4}\,\varDelta\gamma^{1/4}}{E^{\prime1/4}t^{1/4}}.\label{eq:S5:DefinitionQ2D-1}
\end{equation}
Replacing this injection rate into the scales of \citet{List90},
we retrieve exactly the scaling of equation \ref{eq:S5:DefinitionViscHeadScales}.
Rescaled 3D numerical results are shown along with the zero-toughness
solution of \citet{List90} using a tip based coordinate system (${\displaystyle \widehat{z}\left(t\right)=z_{Tip}\left(t\right)-z}$)
in figure~\ref{fig:TipBasedVisc}. The 3D and 2D solutions practically
coincide (relative error of ${\displaystyle \sim5\%}$) for times
${\displaystyle t\gtrsim50t_{m\widehat{m}}}$. In the viscosity-dominated
case, the shrinking of the head indeed reduces the effect of the 3D
curvature at large time (see also the scaled footprint in figure \ref{fig:SelfSimVisc})
and thus renders the elastic state of plane-strain more valid. 

In conclusion, the buoyant viscosity-dominated fracture exhibits a
viscous source region following the \citet{List90b} solution, combined
with a depleting head according to the scaling (\ref{eq:S5:DefinitionViscHeadScales})
at the propagating edge for late times ($t\gg t_{m\widehat{m}}$).
The depleting head follows the solution of a 2D semi-infinite plane-strain
fracture along the centerline. It may be possible to construct a complete
pseudo-3D approximation matching these asymptotes in the source and
head region, a task we leave open for further studies.

\section{Intermediate/finite $\mathcal{M}_{\hat{k}}$ cases}

In the toughness dominated case, we have seen that the $\widehat{\text{K}}$
limit is captured by the \citet{GaGe22} finger-like solution for
${\displaystyle \mathcal{M}_{\widehat{k}}\underset{\sim}{<}1.\times10^{-2}}$.
On the other end, for zero-toughness (${\displaystyle \mathcal{M}_{\widehat{k}}=\infty}$),
horizontal growth continues as $\sim t^{1/4}$ at late times ($t\gtrsim100t_{m\widehat{m}}$).
Numerical results for \emph{large} but finite values of $\mathcal{M}_{\widehat{k}}$
(see insets c and d in figure \ref{fig:ViscFp}) show that, as anticipated
\citep{List90b,GaGe22}, horizontal growth arrests after some time.
The vertical velocity thus increases to a constant terminal velocity
due to volume balance. This is confirmed by the $\mathcal{M}_{\widehat{k}}=500,\,10^{3}$
simulations displayed in figures \ref{fig:ViscFp}c and d (and to
a lesser extent for $\mathcal{M}_{\widehat{k}}=10^{4}$ where the
horizontal arrest was not completely reached). The characteristic timescale
for such a horizontal arrest can be estimated as the time at which
the horizontal dimensionless toughness $\mathcal{K}_{\widehat{m},x}$
(\ref{eq:S5:DefinitionKqx}) in the viscous tail scaling reaches order
one. We obtain
\begin{equation}
\mathcal{K}_{\widehat{m},x}\left(t_{\widehat{m}\widehat{k}}^{x}\right)=1\rightarrow t_{\widehat{m}\widehat{k}}^{x}=\frac{E^{\prime19/5}Q_{o}^{3/5}\mu^{\prime8/5}}{K_{Ic}^{24/5}\varDelta\gamma^{3/5}}=\mathcal{M}_{\widehat{k}}^{36/35}t_{m\widehat{m}}\label{eq:S6_MaxBreadthTime}
\end{equation}
and the corresponding maximum breadth and length scales are
\begin{equation}
b_{\widehat{m}}\left(t_{\widehat{m}\widehat{k}}^{x}\right)=\mathcal{M}_{\widehat{k}}^{2/5}\ell_{b},\hspace*{1em}\ell_{\widehat{m}}\left(t_{\widehat{m}\widehat{k}}^{x}\right)=\ell_{mk}.\label{eq:S6_MaxBreadthScaling-M}
\end{equation}
From our previous discussion, the zero-toughness (${\displaystyle \mathcal{M}_{\widehat{k}}=\infty}$)
self-similar growth is established for $t\gtrsim100t_{m\widehat{m}}$.
For large values of $\mathcal{M}_{\widehat{k}}$, such a zero toughness
solution is thus expected to be realized at intermediate times after
the transition to buoyancy but prior to the characteristic time of
horizontal arrest: for $t\in[100\,t_{m\widehat{m}},t_{\widehat{m}\widehat{k}}^{x}]$.
Using \ref{eq:S6_MaxBreadthTime}, we thus expect to see a period
of lateral growth for dimensionless viscosities at least larger than
$\mathcal{M}_{\widehat{k}}^{36/35}\sim\mathcal{M}_{\widehat{k}}=100$.

\begin{figure}
\centerline{\includegraphics[width=0.9\textwidth]{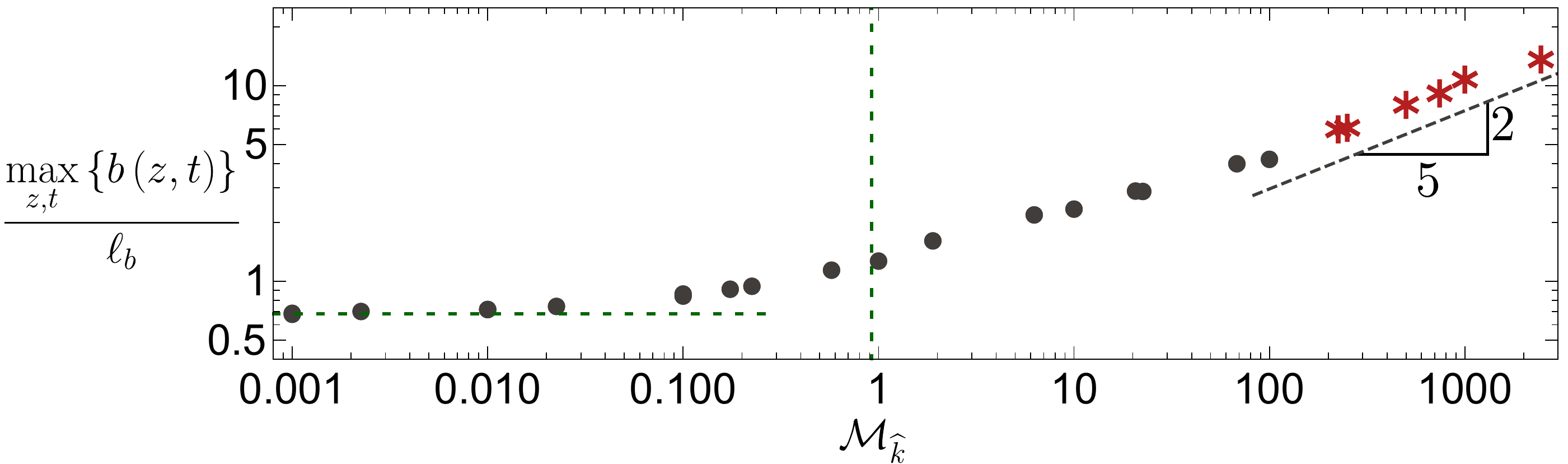}}
\caption{Comparison of maximum breadth for buoyant fractures as a function of
the dimensionless viscosity ${ \mathcal{M}_{\widehat{k}}\in\left[10^{-3},5.\times10^{3}\right]}$.
Black dots are used for fractures with a uniform breadth and red stars
otherwise.The dashed green lines represent the limits of the 3D ${\widehat{\textrm{K}}}$
GG, (2014) solution (${b\sim\pi^{-1/3}\ell_{b}}$
for the breadth limit (horizontal line) and ${ \mathcal{M}_{\widehat{k}}\approx0.92}$
for the stabilization criterion (vertical line)). The grey dashed line emphasizes the scaling relation
${ \protect\underset{z,t}{\textrm{max}}\left\{ b\left(z,t\right)\right\} \sim\mathcal{M}_{\widehat{k}}^{2/5}\ell_{b}}$.\label{fig:BreadthIntermediate}}
\end{figure}

We performed a series of simulations spanning a wide range of values
of $\mathcal{M}_{\widehat{k}}$ from $10^{-3}$ to $10^{3}$ for which
the simulations were run long enough to observe a cessation of horizontal
growth. We report in figure \ref{fig:BreadthIntermediate} the evolution
of the maximum breadth of the buoyant fracture with $\mathcal{M}_{\widehat{k}}$.
As expected, in the toughness dominated limit $\mathcal{M}_{\widehat{k}}<1$,
the fracture breadth remains close to the $\widehat{\text{K}}$ limit.
The maximum breadth then increases with $\mathcal{M}_{\widehat{k}}$
from the \citet{GaGe22} $b\sim\pi^{-1/3}\ell_{b}$ solution for $\mathcal{M}_{\widehat{k}}<10^{-2}$
up to $b\sim5\ell_{b}$ for $\mathcal{M}_{\widehat{k}}=100$. For
values up to $\mathcal{M}_{\widehat{k}}\sim100$, we always observe
a uniform breadth along the fracture footprint and no horizontal growth
is observed after the transition to buoyancy. These fractures have
a clear finger-like shape. It is worth noting that from their approximate
3D toughness solution \citet{GaGe22} obtain a lower value ${\displaystyle \mathcal{M}_{\widehat{k}}\approx0.92}$
as a criterion for no further horizontal growth. Accounting for fully
3D effects, the domain of ``finger-like'' fracture shapes is seen
to extend up to $\mathcal{M}_{\widehat{k}}=100$.

\begin{figure}
\centerline{\includegraphics[width=0.75\textwidth]{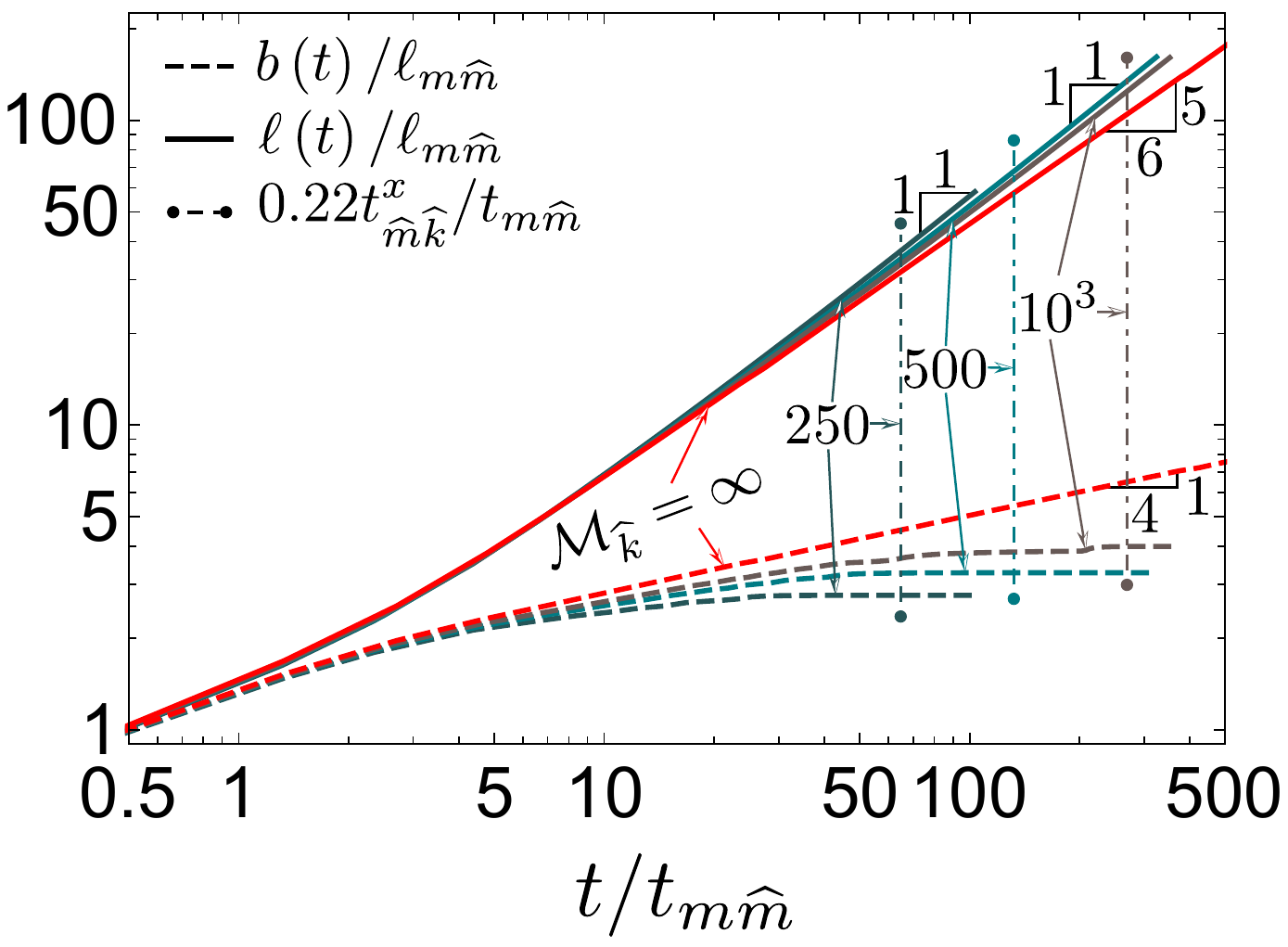}}
\caption{Evolution of fracture breadth and length for intermediate fractures
without a uniform breadth ${ \mathcal{M}_{\widehat{k}}\in[10^{2}-2\times10^{3}]}$
(the simulation with ${ \mathcal{M}_{\widehat{k}}=\infty}$
is used as a reference). Dashed lines show fracture breadth, continuous
lines fracture height, and horizontal dashed-dotted lines the expected
time where lateral growth stops. We indicate the emerging power laws
on the figure.\label{fig:TimeEvolutionIntermediate}}
\end{figure}

For values of $\mathcal{M}_{\widehat{k}}>100$, the fractures have
a distinctly different late time shape akin to an inverted cudgel (non-uniform horizontal breadth) with an ultimately fixed maximum
horizontal breadth. We recover the predicted evolution of the maximum
breadth \ref{eq:S6_MaxBreadthScaling-M} as ${\displaystyle \mathcal{M}_{\widehat{k}}^{2/5}\ell_{b}}$
(red stars in figure \ref{fig:BreadthIntermediate}). A fit of our
numerical results actually provides ${\displaystyle \underset{z,t}{\textrm{max}}\left\{ b\left(z,t\right)\right\} \approx0.6858\mathcal{M}_{\widehat{k}}^{2/5}\ell_{b}}$
for ${\displaystyle \mathcal{M}_{\widehat{k}}\in[10^{2}-2\times10^{3}]}$.
Using this fitted pre-factor on the breadth evolution, assuming $b\sim b_{\widehat{m}}(t)$
before stabilization, we estimate the time for breadth stabilization
to be ${\displaystyle \sim0.22t_{\widehat{m}\widehat{k}}^{x}}$. We
graphically show in figure \ref{fig:TimeEvolutionIntermediate} that
this estimation agrees fairly well with the numerical results.
For the reported simulations, the fracture length
ultimately evolves linearly in time (indicated by a 1 to 1 slope in
figure \ref{fig:TimeEvolutionIntermediate}) as ${\displaystyle \ell\left(t\right)/\ell_{m\widehat{m}}\sim\mathcal{M}_{\widehat{k}}^{-6/35}\left(t/t_{m\widehat{m}}\right)}$.

We also performed simulations for $\mathcal{M}_{\widehat{k}}>10^{3}$
, which however did not reach the arrest of horizontal growth within
a reasonable computational time limit. It is worth pointing out that
from these numerical results, the self-similar viscous (${\displaystyle \mathcal{M}_{\widehat{k}}=\infty}$)
evolution is actually visible at intermediate times only for dimensionless
viscosities larger than $10^{4}$ (see figures \ref{fig:ViscFp}c and
d).

\section{Discussion\label{sec:Discussion}}

\subsection{Orders of magnitude}

In nature, buoyant hydraulic fractures are suggested to be a major
contributor to the transport of magma through the lithosphere \citep{RiTa15}.
For such cases, data collection is difficult and often restricted
to the investigation of outcrops from dikes. A broad range of rarely
well-constrained parameters is possible. We thus only briefly illustrate
the emergence of dikes using the following parameters \citep{MoLe21}:
${\displaystyle E^{\prime}\sim10\,\textrm{GPa}}$, ${\displaystyle K_{Ic}\sim1.5\,\textrm{MPa}\cdot\textrm{m}^{1/2}}$,
${\displaystyle \mu_{f}=100\,\textrm{Pa}\cdot\textrm{s}}$,
${\displaystyle \varDelta\rho\sim250\,\textrm{kg}\cdot\textrm{m}^{-3}}$,
and a low value of the release rate ${\displaystyle Q_{o}\sim1\,\textrm{m}^{3}\cdot\textrm{s}^{-1}}$.
For this set of parameters, the dike intrusion is strongly viscosity
dominated with ${\displaystyle \mathcal{M}_{\widehat{k}}\approx3.29\times10^{6}}$
and has a maximum lateral extent of tens of kilometres. The use of
a higher release rate would linearly increase the value of ${\displaystyle \mathcal{M}_{\widehat{k}}}$
and thus only render the growth more viscous dominated. The corresponding
fracture height easily exceeds the thickness of the lithosphere, as
already pointed out by \citet{List90b}. As a result, such large extents
will necessarily clash with the length scales of stress and material
heterogeneities. It also indicates the very strong effect of buoyancy
on upward growth. 

\subsection{Comparison with experiments}

\begin{figure}
\begin{centering}
\includegraphics[width=0.9\textwidth]{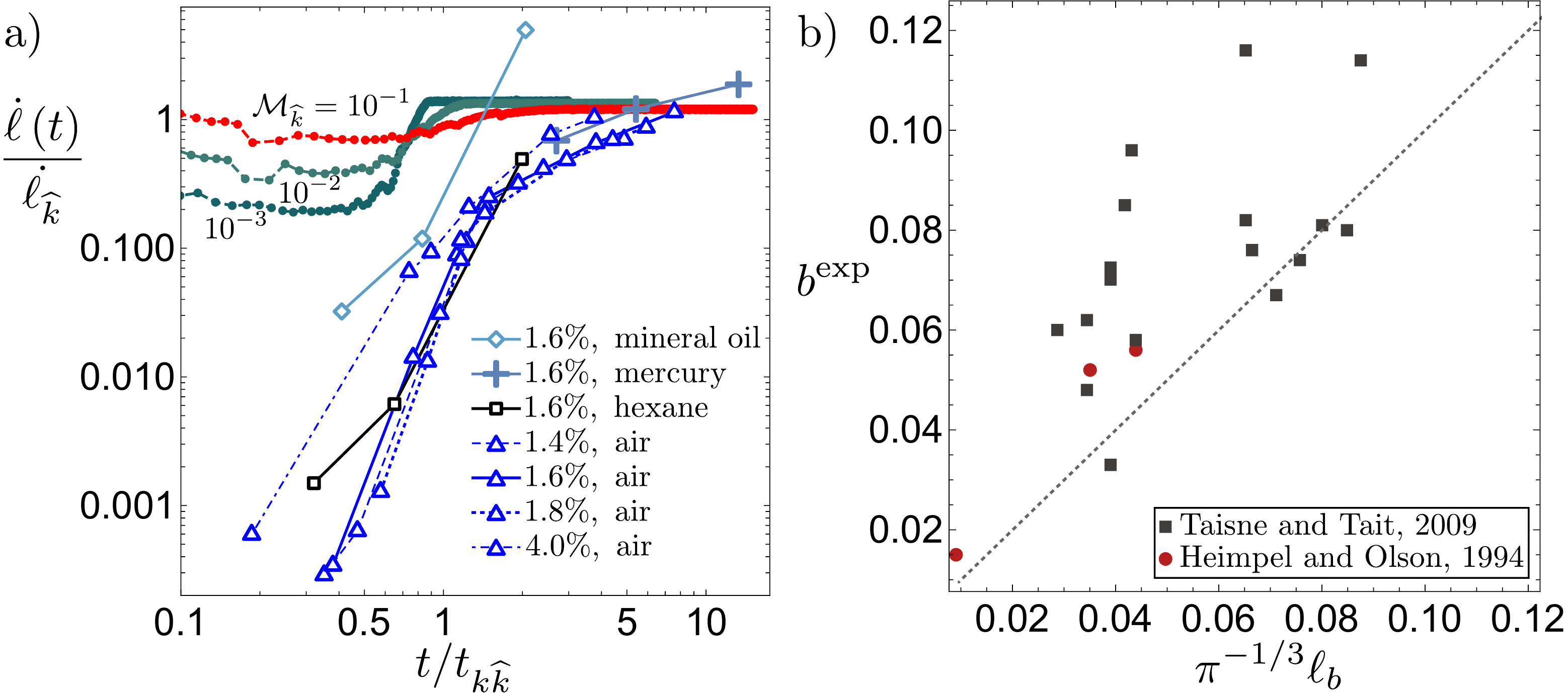}
\par\end{centering}
\caption{a) Comparison of the experiments of \citet{HeOl94} with our simulations.
The experiment takes place within the transient and the initiation
already favors the buoyant propagation. b) Comparison of estimated
and observed breadth for various experimental studies.\label{fig:ExperimentComparison}}
\end{figure}

Various experiments on buoyant fractures have been performed in the laboratory
\citep{HeOl94,TaTa09,RiBo05,TaTa11,ItMa02}. Most of these experiments
consist of a finite (not continuous) release and aim at investigating
various mechanisms (arrest due to material heterogeneities among others).
We evaluate in figure \ref{fig:ExperimentComparison}a the evolution
of the fracture velocity with time for the experiments performed by
\citet{HeOl94}. The data in their figure 2 is transformed to correspond
to our scaled velocity and time. All experimental parameters except the release rate $Q_o$ are taken from \citet{HeOl94}. The good match of figure \ref{fig:ExperimentComparison}a was obtained using an estimate of the release rate of ${\displaystyle Q_{o}\sim10^{-8}\,\textrm{m}^{3}\cdot\textrm{s}^{-1}}$. The corresponding dimensionless
viscosities range between ${\displaystyle \mathcal{M}_{\widehat{k}}\in\left[8.8\times10^{-8},2.3\times10^{-3}\right]}$
(see details in supplementary material). When superimposing their
velocity evolution with our numerical results for ${\displaystyle \mathcal{M}_{\widehat{k}}\in\left[10^{-3},10^{-1}\right]}$,
we observe that their experiments start in the transition between
the radial and buoyant regimes. In other words, their experiments
are situated within the accelerating phase and their velocities tend
to stabilize only towards the very end of the experiment. Some experiments
show a deceleration but do not quite reach a constant velocity as
the time to overcome the transient (${\displaystyle t/t_{k\widehat{k}}\approx14})$
is reached in none of the experiments. This is a direct consequence
of the limited sample size, which is insufficient in all experiments
to reach the end of the transient regime (see details in supplementary material).
We thus conclude that these experiments are strongly influenced by
their initial conditions (a too large initial notch) and the finiteness
of the specimens which prevents them from reaching the constant terminal
velocity.

As described in section \ref{subsec:S4:TransitionToughness}, in that
range of such low dimensionless viscosities, we can nevertheless compare
the fracture breadth in the transient phase. We could extract information
on the fracture breadth from two contributions albeit with uncertainties
on some reported parameters. We assume that for such toughness-dominated
buoyant fractures, the $\widehat{\text{K}}$ solution of \citet{GaGe22}
is also valid in the case of a finite volume release, which allows
us to use the data from \citet{TaTa09}. We report in figure \ref{fig:ExperimentComparison}b
the measured breadth ${\displaystyle b^{\textrm{exp}}}$ and compare
it to the limiting 3D $\widehat{\text{K}}$ GG, (2014) solution of ${\displaystyle \pi^{-1/3}\ell_{b}}$.
The breadth is generally underestimated for both contributions.
In most cases, the extension of the fracture in these experiments
clashes with the finite size of the sample, and the initial notch size might be inadequate. These boundary and initiation effects
may also modify the linear gradient of the background stress and thus
render the evaluation of ${\displaystyle \varDelta\gamma}$ erroneous.

\subsection{Possibility of approximate solutions}

The computational cost of the reported simulations is considerable and tests the limits of the numerical solver used herein (see section \ref{subsec:S2:NumSolv} for details). For example, the simulations presented in figures \ref{fig:ToughFp} and \ref{fig:ViscFp} took between two to two and a half weeks on a multithreaded Linux desktop system with twelve Intel\textregistered Core i7-8700 CPUs and used at most 30 GB of RAM. Such requirements are common for the simulations presented in this contribution. 

Interestingly, our results point to the possible development of reduced-order pseudo-3D models \citep{AdPe08,AdDe10} that would inevitably be much more computationally efficient. For example, the 3D  $\widehat{\text{K}}$ GG, (2014) solution of \citet{GaGe22} is based on a finger-like fracture approximation for the tail while keeping a complete description of the elasticity in the head region. We could demonstrate the validity of this assumption as discussed in section \ref{sec:ToughnessDominated}.
Employing the knowledge gained from our results, the development of accurate and computationally efficient models similar to the ones presented in \citet{DoPe15} may be possible. The solution derived in \citet{List90b} is based on a similar approach for the zero-toughness case. We could show that this approach works fairly well within the source region but fails to capture the transition to the head region, which has not been prescribed in the work of \citet{List90b}. The insights gained from our simulations (see section \ref{subsec:ViscSS}) could be used to further develop an enhanced pseudo-3D model for the viscous case. Such a model could then possibly bridge the source-region solution of \citet{List90b} with a viscous head. 

\section{Conclusions}

\begin{figure}
\centerline{\includegraphics[width=0.9\textwidth]{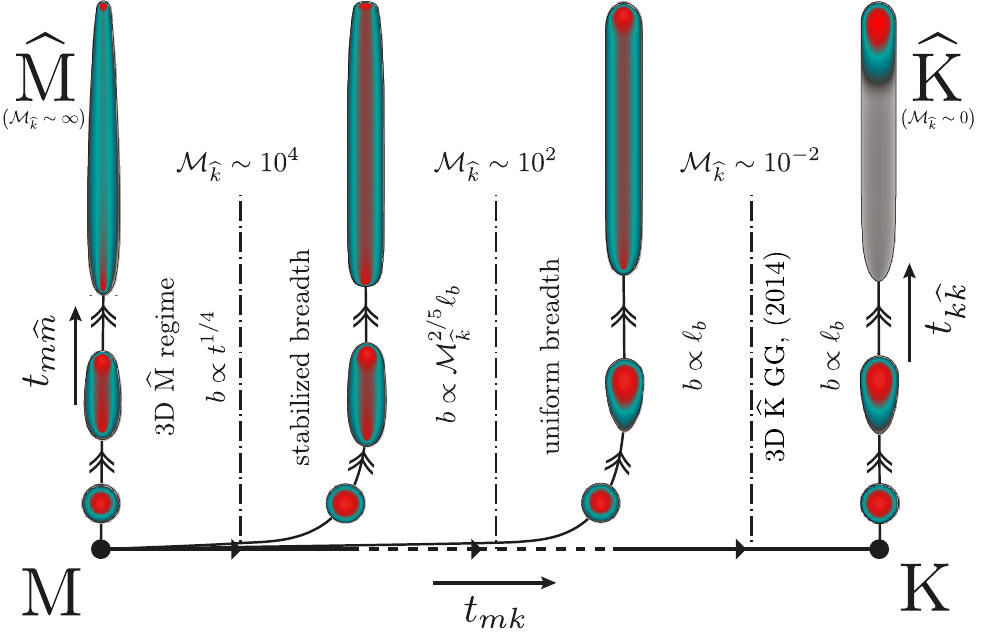}}
\caption{Propagation diagram for 3D buoyant fractures under a continuous fluid
release. Radial growth is initially viscosity-dominated (${  \textrm{M}}$-vertex).
Transition to buoyancy either occurs before (${  \mathcal{M}_{\widehat{k}}}\gg1$)
or after (${  \mathcal{M}_{\widehat{k}}}\gg1$) the transition
to radial toughness-dominated growth. At late times, a family of buoyancy-driven
solutions as a function of ${  \mathcal{M}_{\widehat{k}}}$
(\ref{eq:S3:DefinitionMkhat}) emerges. The large toughness limit
(section \ref{sec:ToughnessDominated}) is reached for values of ${  \mathcal{M}_{\widehat{k}}\protect\underset{\sim}{<}10^{-2}}$,
whereas the zero-toughness solution (section \ref{sec:ViscDominate})
appears at intermediate times $t\in [100t_{m\widehat{m}},t_{\widehat{m}\widehat{k}}^{x}] $
for ${ \mathcal{M}_{\widehat{k}}\protect\underset{\sim}{>}10^{4}}$.
\label{fig:ParametricSpace}}
\end{figure}
 For a homogeneous linear elastic solid subjected to a linear background
confining stress and a Newtonian fluid, using numerical simulations
and scaling analysis, we have shown that under a constant release
rate the growth of 3D buoyant fractures is governed by a single dimensionless
number $\mathcal{M}_{\widehat{k}}$ (\ref{eq:S3:DefinitionMkhat}).
It is worth emphasizing the very large computational cost of the simulations
reported here which span more than ten, respectively twenty, orders
of magnitude in space and time. They reach the computational limit
of our current implementation of the implicit level set algorithm. Nonetheless, from this series of simulations we have shown that a
family of buoyant HF emerge at late times as a function of $\mathcal{M}_{\widehat{k}}$ (HF = Hydraulic Fractures – see def at the beginning of the paper).
The solution phase space can be summarized in the diagram displayed
in figure \ref{fig:ParametricSpace}. At early time, all fractures
start with a radial shape and are initially dominated by viscous dissipation
(${\displaystyle \textrm{M}}$-vertex), and remain radial for times
lower than the buoyancy transition time scales \ref{eq:S3:DefinitionTmmAndTkk}.
Depending on the ratio between the radial viscosity to toughness transition
time-scale $t_{mk}$ (without buoyancy) and the viscous buoyancy transition
time-scale $t_{m\widehat{m}}$ (or $t_{k\widehat{k}}$), encapsulated
in the definition of the dimensionless viscosity $\mathcal{M}_{\widehat{k}}$
(\ref{eq:S3:DefinitionMkhat}), a family of solutions exists at late
time when buoyancy dominates. If the transition to buoyancy occurs
when the hydraulic fracture is already in the toughness-dominated
regime (${\displaystyle \mathcal{M}_{\widehat{k}}\underset{\sim}{<}10^{-2}}$),
the late time growth is well captured by the ${\displaystyle \widehat{\textrm{K}}}$
approximate solution of \citet{GaGe22}. In this limit of large toughness,
the buoyant HF has a distinct toughness-dominated head with a constant
volume and shape, and a viscosity-dominated tail that governs its
upward growth. For an intermediate range of ${\displaystyle \mathcal{M}_{\widehat{k}}\in\left[10^{-2},10^{2}\right]}$,
the fracture remains finger-like with a uniform breadth for each cross-section
albeit with an increasing breadth with $\mathcal{M}_{\widehat{k}}$.
Above $\mathcal{M}_{\widehat{k}}>100$, the hydraulic fractures exhibit
an inverted cudgel shape at late time (the breadth is no longer spatially
uniform in the tail) and the maximum horizontal breadth increases
as $\mathcal{M}_{\widehat{k}}^{2/5}\ell_{b}$ as horizontal growth
occurs until a given time $t_{\widehat{m}\widehat{k}}^{x}$ (\ref{eq:S6_MaxBreadthTime}).
For values of ${\displaystyle \mathcal{M}_{\widehat{k}}\underset{\sim}{>}10^{4}}$,
a zero-toughness self-similar $\widehat{\text{M}}$ limit (section
\ref{sec:ViscDominate}) can be observed at intermediate times. This
self-similar $\widehat{\text{M}}$ viscosity-dominated limit exhibits
an ever-increasing breadth in association with the zero toughness
assumption. The scaling of the $\widehat{\text{M}}$, regime originally
presented in \citet{List90b}, is confirmed by our numerical results.
In that limit, the viscous head is slowly depleting with time with
a centerline evolution akin to the known 2D plane-strain near-tip
asymptotic solution at late time. It might be possible to develop
an approximate solution for that viscous limit along similar lines
as in the toughness-dominated case when combining the source solution
and the near-tip viscous head. A finite toughness always ensures an
ultimate arrest of horizontal growth at a characteristic time $t_{\widehat{m}\widehat{k}}^{x}=\mathcal{M}_{\widehat{k}}^{36/35}t_{m\widehat{m}}$
for which the horizontal dimensionless toughness becomes of order
one. Besides their final shapes, another important difference between
buoyant toughness dominated HF and viscous ones lie in the transition
to the buoyant regime. For toughness-dominated fractures, a significant
vertical acceleration ($\propto\mathcal{M}_{\widehat{k}}^{-1/3}$)
is observed whereas viscosity-dominated fractures have a smoother
vertical acceleration thanks to horizontal growth.

Natural magmatic buoyant fractures are likely always viscosity-dominated,
while on the other hand all laboratory experiments have been performed
under toughness dominated conditions. It appears that even in the
toughness regime, precise experiments are still lacking for quantitative
comparison with the theoretical predictions reported here for buoyant
fractures. Orders of magnitude for magmatic dikes also indicate that
their horizontal and vertical extent will necessarily clash with length
scales of stress and material heterogeneities at late times. These
heterogeneities, as well as the possibility of fluid exchange with
the surrounding rock and thermal effects, may play a critical role
in the growth and potential arrest of buoyant hydraulic fractures
on their way towards the surface. The interplay of these effects on
linear hydraulic fracture mechanics growth remains to be investigated. Finally, most fluid releases are of a finite volume rather than having an ever-ongoing release at a constant injection rate. This particular problem is part of ongoing research and is essentially based on the findings presented in this contribution.

\vspace{0.25cm}

\noindent \textbf{Acknowledgements.} The authors gratefully acknowledge
in-depth discussions with Dmitry Garagash regarding the 3D toughness
dominated approximate solution presented in \citet{GaGe14,GeGa14,GaGe22}.
\\
\textbf{Funding.} This work was funded by the Swiss National Science
Foundation under grant \#192237. \\
\textbf{Declaration of Interest. }The authors report no conflict of
interest.\\
\textbf{Data availability statement.} 
The version of the open-source solver PyFrac, corresponding scripts and results of simulation
of this study are openly available at \href{http://doi.org/10.5281/zenodo.6511166}{10.5281/zenodo.6511166}.
\\
\textbf{Author ORCID. }A. M\"ori, \href{https://orcid.org/0000-0002-7951-1238}{0000-0002-7951-1238};
B. Lecampion, \href{https://orcid.org/0000-0001-9201-6592}{0000-0001-9201-6592}\\
\textbf{Author contributions. }Andreas M\"ori: Conceptualization, Methodology,
Formal analysis, Investigation, Software, Validation, Visualization,
Writing - original draft. Brice Lecampion: Conceptualization, Methodology,
Formal analysis, Validation, Supervision, Funding acquisition, Writing
- review \& editing.

\appendix

\section{Recapitulating tables of scales\label{Appsec:ScalingsTables}}

For completeness, we list all the scales used within this contribution
in the following tables. A Wolfram mathematica notebook containing
their derivation and the different scalings is further provided in
a supplementary material.

\begin{table}
\begin{center}
\def~{\hphantom{0}}
\begin{scriptsize}
\begin{tabular}{ccccccc}
 & \multicolumn{2}{c}{\emph{radial}} & \multicolumn{4}{c}{\emph{elongated}}\\[3pt]

 & $\text{M}$ & $\text{K}$ & $\widehat{\text{M}}$ (tail) & $\widehat{\text{M}}$ (head) & $\widehat{\text{K}}$ (tail) & $\widehat{\text{K}}$ (head)\\[3pt]
$\ell_{*}$ & $\dfrac{E^{\prime1/9}Q_{o}^{1/3}t^{4/9}}{\mu^{\prime1/9}}$ & $\dfrac{E^{\prime2/5}Q_{o}^{2/5}t^{2/5}}{K_{Ic}^{2/5}}$ & $\dfrac{Q_{o}^{1/2}\varDelta\gamma^{1/2}t^{5/6}}{E^{\prime1/6}\mu^{\prime1/3}}$ & $\dfrac{E^{\prime11/24}Q_{o}^{1/8}\mu^{\prime1/6}}{\varDelta\gamma^{5/8}t^{1/24}}$ & $\dfrac{Q_{o}^{2/3}\varDelta\gamma^{7/9}t}{K_{Ic}^{4/6}\mu^{\prime1/3}}$ & $\ell_{b}$\\[2pt]

$b_{*}$ & $\ell_{*}$ & $\ell_{*}$ & $\dfrac{E^{\prime1/4}Q_{o}^{1/4}t^{1/4}}{\varDelta\gamma^{1/4}}$ & $\ell_{*}$ & $\ell_{b}=\dfrac{K_{Ic}^{2/3}}{\varDelta\gamma^{2/3}}$ & $\ell_{*}$\\[2pt]

$w_{*}$ & $\dfrac{Q_{o}^{1/3}\mu^{\prime2/9}t^{1/9}}{E^{\prime2/9}}$ & $\dfrac{K_{Ic}^{4/5}Q_{o}^{1/5}t^{1/5}}{E^{\prime4/5}}$ & $\dfrac{Q_{o}^{1/4}\mu^{\prime1/3}}{E^{\prime1/12}\varDelta\gamma^{1/4}t^{1/12}}$ & $w_{*}^{\text{tail}}$ & $\dfrac{Q_{o}^{1/3}\mu^{\prime1/3}}{K_{Ic}^{2/9}\varDelta\gamma^{1/9}}$ & $w_{k\widehat{k}}$\\[2pt]

$V_{*}$ & $Q_{o}t$ & $Q_{o}t$ & $Q_{o}t-V_{*}^{\text{head}}$ & $\dfrac{E^{\prime5/6}Q_{o}^{1/2}\mu^{\prime2/3}}{\varDelta\gamma^{3/2}t^{1/6}}$ & $Q_{o}t-V_{*}^{\text{head}}$ & $\dfrac{K_{Ic}^{8/3}}{E^{\prime}\varDelta\gamma^{5/3}}$\\

$p_{*}$ & $\dfrac{E^{\prime2/3}\mu^{\prime1/3}}{t^{1/3}}$ & $\dfrac{K_{Ic}^{6/5}}{E^{\prime1/5}Q_{o}^{1/5}t^{1/5}}$ & $\dfrac{E^{\prime2/3}\mu^{\prime1/3}}{t^{1/3}}$ & $\dfrac{E^{\prime11/24}Q_{o}^{1/8}\varDelta\gamma^{3/8}\mu^{\prime1/6}}{t^{1/24}}$ & $\dfrac{E^{\prime}\varDelta\gamma^{5/9}Q_{o}^{1/3}\mu^{\prime1/3}}{K_{Ic}^{8/9}}$ & $p_{k\widehat{k}}$\\[2pt]

\multirow{2}{*}{$\mathcal{P}_{s}$} & $\mathcal{K}_{m}=(t/t_{mk})^{1/9}$ & $\mathcal{M}_{k}=(t/t_{mk})^{-2/5}$ & \multicolumn{2}{c}{\multirow{2}{*}{$\mathcal{K}_{\widehat{m},x}=\mathcal{M}_{\widehat{k}}^{-3/14}\left(t/t_{m\widehat{m}}\right)^{5/24}$}} & \multicolumn{2}{c}{\multirow{2}{*}{$\mathcal{M}_{\widehat{k}}=\mu^{\prime}\dfrac{Q_{o}E^{\prime3}\varDelta\gamma^{2/3}}{K_{Ic}^{14/3}}$}}\\[2pt]

& $\mathcal{B}_{m}=(t/t_{m\widehat{m}})^{7/9}$ & $\mathcal{B}_{k}=(t/t_{k\widehat{k}})^{3/5}$ & &\\
\end{tabular}
\end{scriptsize}
\caption{Characteristic scales (and governing dimensionless parameters $\mathcal{P}_{s}$) in
the different scalings.}
\label{Apptab:RadialScales}
\end{center}
\end{table}

\begin{table}
\begin{center}
\def~{\hphantom{0}}
\begin{scriptsize}
\begin{tabular}{ccccc}
 & $t$ & $\ell_{*}=b_{*}$ & $w_{*}$ & $p_{*}$\\[3pt]

$\text{M}\rightarrow\text{K}$ & $t_{mk}=\dfrac{E^{\prime13/2}Q_{o}^{3/2}\mu^{\prime5/2}}{K_{Ic}^{9}}$ & $\ell_{mk}=\dfrac{E^{\prime3}Q_{o}\mu^{\prime}}{K_{Ic}^{4}}$ & $w_{mk}=\dfrac{E^{\prime1/2}Q_{o}^{1/2}\mu^{\prime1/2}}{K_{Ic}}$ & $p_{mk}=\dfrac{K_{Ic}^{3}}{E^{\prime3/2}Q_{o}^{1/2}\mu^{\prime1/2}}$\\[2pt]

$\text{M}\rightarrow\widehat{\text{M}}$ & $t_{m\widehat{m}}=\dfrac{E^{\prime5/7}\mu^{\prime4/7}}{Q_{o}^{3/7}\varDelta\gamma^{9/7}}$ & $\ell_{m\widehat{m}}=\dfrac{E^{\prime3/7}Q_{o}^{1/7}\mu^{\prime1/7}}{\varDelta\gamma^{4/7}}$ & $w_{m\widehat{m}}=\dfrac{Q_{o}^{2/7}\mu^{\prime2/7}}{E^{\prime1/7}\varDelta\gamma^{1/7}}$ & $p_{m\widehat{m}}=E^{\prime3/7}Q_{o}^{1/7}\mu^{\prime1/7}\varDelta\gamma^{3/7}$\\[2pt]

$\text{K}\rightarrow\text{\ensuremath{\widehat{\text{K}}}}$ & $t_{k\widehat{k}}=\dfrac{K_{Ic}^{8/3}}{E^{\prime}Q_{o}\varDelta\gamma^{5/3}}$ & $\ell_{k\widehat{k}}=\ell_{b}=\dfrac{K_{Ic}^{2/3}}{\varDelta\gamma^{2/3}}$ & $w_{k\widehat{k}}=\dfrac{K_{Ic}^{4/3}}{E^{\prime}\varDelta\gamma^{1/3}}$ & $p_{k\widehat{k}}=K_{Ic}^{2/3}\varDelta\gamma^{1/3}$\\[2pt]
\end{tabular}
\end{scriptsize}
\caption{Transition scales between regimes. The toughness head scales in table \ref{Apptab:RadialScales}
corresponds to the transition scales $\text{K}\rightarrow\text{\ensuremath{\widehat{\text{K}}}}$.}
\label{Apptab:TransitionScales}
\end{center}
\end{table}

%\pagebreak{}

\bibliographystyle{jfm}
\bibliography{GeneralBiblio}

\end{document}